\documentclass[headings=standardclasses]{scrartcl}

\usepackage{geometry}
\usepackage[utf8]{inputenc}
\usepackage[T1]{fontenc}
\usepackage[onehalfspacing]{setspace}
\usepackage{times}
\usepackage[charter]{mathdesign}
\usepackage{microtype}
\usepackage{amsmath}
\usepackage{statmath}
\usepackage{amsthm}
\usepackage{tabulary}
\usepackage{longtable}
\usepackage{booktabs} 
\usepackage{diagbox}
\usepackage{ragged2e}

\usepackage{siunitx}
\usepackage{bm}
\usepackage[pdftex]{graphicx}
\usepackage{rotating}
\usepackage{pdfpages}
\usepackage{pdflscape}
\usepackage{color}
\usepackage{array}
\usepackage[font=bf]{caption}
\usepackage{subcaption}
\usepackage{pgfplots}
\usepackage{pgfplotstable}
\usepackage{float}
\usepackage[title]{appendix}
\pgfplotsset{compat=1.9}
\usepackage{url}
\usepackage[section]{placeins}
\usepackage{natbib}
\usepackage{hyperref}
\hypersetup{
	colorlinks   = true,    
	urlcolor     = blue,    
	linkcolor    = blue,    
	citecolor    = blue      
}
\usepackage{pifont}

\makeatletter
\renewcommand*\env@matrix[1][c]{\hskip -\arraycolsep
  \let\@ifnextchar\new@ifnextchar
  \array{*\c@MaxMatrixCols #1}}
\makeatother

\usepackage{xcolor}
\usepackage{soul}

\usepackage{verbatim}

\begin{document}
\thispagestyle{empty}
\begin{spacing}{1.2}
\begin{flushleft}
\huge \textbf{Revisiting McFadden's correction factor for sampling of alternatives in multinomial logit and mixed multinomial logit models}\\
\vspace{\baselineskip}
\normalsize
29 September 2023 \\
\vspace{\baselineskip}
\textsc{Thijs Dekker} (corresponding author) \\
Institute for Transport Studies\\
University of Leeds, UK \\
t.dekker@leeds.ac.uk \\
\vspace{\baselineskip}
\textsc{Prateek Bansal} \\
Department of Civil and Environmental Engineering\\
National University of Singapore \\
prateekb@nus.edu.sg \\
\vspace{\baselineskip}
\textsc{Jinghai Huo} \\
Department of Civil and Environmental Engineering\\
National University of Singapore \\
jinghaihuo@u.nus.edu \\
\end{flushleft}
\end{spacing}

\newpage
\thispagestyle{empty}
\section*{Abstract}
In this paper, we revisit \citet{mcfadden1978modeling}'s correction factor for sampling of alternatives in multinomial logit (MNL) and mixed multinomial logit (MMNL) models. \citet{mcfadden1978modeling} proved that consistent parameter estimates are obtained when estimating MNL models using a sampled subset of alternatives, including the chosen alternative, in combination with a correction factor. We decompose this correction factor into i) a correction for overestimating the MNL choice probability due to using a smaller subset of alternatives, and ii) a correction for which a subset of alternatives is contrasted through utility differences and thereby the extent to which we learn about the parameters of interest in MNL. \citet{keane2012estimation} proved that the overall expected positive information divergence - comprising the above two elements - is minimised between the true and sampled likelihood when applying a sampling protocol satisfying uniform conditioning. We generalise their result to the case of positive conditioning and show that whilst \citet{mcfadden1978modeling}'s correction factor may not minimise the overall expected information divergence, it does minimise the expected information loss with respect to the parameters of interest. We apply this result in the context of Bayesian analysis and show that \citet{mcfadden1978modeling}'s correction factor minimises the expected information loss regarding the parameters of interest across the entire posterior density irrespective of sample size. In other words, \citet{mcfadden1978modeling}'s correction factor has desirable small and large sample properties. The joint application of Bayesian estimation and sampling of alternatives could be attractive for computationally expensive MMNL models with large choice sets. Specifically, we show that our results for Bayesian MNL models transfer to MMNL and that only \citet{mcfadden1978modeling} correction factor is sufficient to minimise the expected information loss in the parameters of interest. Namely, the implementation of data augmentation in Bayesian estimation overcomes the challenges set out by \cite{guevara2013sampling} and \cite{keane2012estimation} in defining feasible correction factors for sampling of alternatives under MMNL using classical estimation. Monte Carlo simulations illustrate the successful application of sampling of alternatives in Bayesian MMNL models. 

\bigskip

\emph{Keywords:} Multinomial logit, Mixed Multinomial logit, Sampling of alternatives, information loss, Bayesian estimation 

\newpage
\pagenumbering{arabic}

\section{Introduction}
\noindent Recent works in transportation, environmental economics and marketing \citep{Dalyetal2014,guevara2013sampling,Guevara2013,keane2012estimation,SINHA2018,Tsoleridis2021,von2018estimation} have renewed interest in the sampling of alternatives in the context of discrete choice modelling. Sampling of alternatives reduces the computational challenge of evaluating the denominator of the logit choice probability for large choice sets by only making use of a smaller subset of sampled alternatives including the chosen alternative. The benefit of sampling alternatives stems from a significant reduction in the estimation time of large-scale choice models. Evaluating the logit formula over a subset of alternatives by default overestimates the choice probability which in turn may have unintended consequences for estimating model parameters. \citet{mcfadden1978modeling} already proved for multinomial logit (MNL) models that an application of a sampling correction to the utility function of the sampled alternatives results in consistent parameter estimates. 

Most discrete choice applications nowadays apply model specifications beyond MNL. Models from the Multivariate Extreme Value (MEV) family, such as the nested logit model \citep{Daly1987}, and mixed multinomial logit (MMNL) models \citep{ReveltTrain1998} have become state of the art. \cite{Guevara2013} and \cite{guevara2013sampling} have proved that \citet{mcfadden1978modeling}'s proposed correction factor forms only a part of the solution to implement sampling of alternatives within these more advanced models. Consistent parameter estimates can be obtained when in addition to McFadden's correction term, the analyst also corrects for the imperfect representation of the LogSum in MEV in the sampled model and the latent nature of individual-level parameters in MMNL.

\citet{guevara2013sampling} highlight in their conclusions the need to study sampling of alternatives in the context of finite sample sizes and alternative estimation strategies. \citet{von2018estimation} argue consistency of parameter estimates and welfare measures of latent class models using the Expectation-Maximisation (EM) algorithm in combination with sampling of alternatives. They also empirically show good performance for relatively small sizes of sampled choice sets, up to 5\% of the full choice set size, but do not address the issue of small sample sizes. In this paper, we study the sampling of alternatives in the context of Bayesian estimation routines for MNL and MMNL models. Since the Bayesian approach does not rely on asymptotically large sample sizes, our paper addresses both aspects highlighted by \citet{guevara2013sampling}.

With the advancements in computational resources and approximate inference, Bayesian estimation has gained popularity in the discrete choice modelling literature \citep{BANSAL2020}. Bayesian estimation is particularly fruitful when latent constructs are included in the likelihood function. Significant reductions in estimation time over classical maximum simulated likelihood (MSL) approaches are typically obtained due to augmenting latent constructs \citep{Tanner1987,Train2009}, such as individual-level parameters in mixed logit models or class membership in latent class models. The joint implementation of sampling of alternatives and Bayesian estimation may therefore lead to additional time savings and spur the implementation of more flexible discrete choice models in travel demand models with large choice sets.

We primarily take a theoretical approach and revisit \citet{mcfadden1978modeling}'s correction factor in the context of MNL in Section \ref{Sec:MNL}. We provide an intuitive explanation of the role of the correction factor in the numerator and denominator of the corrected MNL choice probability under the sampling of alternatives. We decompose \citet{mcfadden1978modeling}'s correction factor into i) a correction for overestimating the MNL choice probability due to using a smaller subset of alternatives, and ii) a correction for which subset of alternatives is contrasted through utility differences and thereby the extent to which we learn about the parameters of interest in MNL. Building on the work of \citet{keane2012estimation}, we show that these two components comprise the loss of information (or information divergence) between the 'true' and 'sampled' MNL log-likelihood. We show that only the second component - operating through the denominator of the MNL choice probability - is relevant for estimation purposes. According to our results, \citet{mcfadden1978modeling}'s correction factor not only results in consistent parameter estimates but also minimises the expected loss of information with respect to the parameters of interest in MNL. This result generalises and puts context to \citet{keane2012estimation}'s conclusions on the expected information divergence under the sampling of alternatives in MNL.     

The outcome that \citet{mcfadden1978modeling}'s correction factor minimises the expected loss of information with respect to the parameters of interest in MNL is central to implementing sampling of alternatives in Bayesian MNL and MMNL models (and thereby establishing the desirable small sample properties of \citet{mcfadden1978modeling}'s correction factor) in Sections \ref{Sec:BayesMNL} and \ref{Sec:BayesMMNL}, respectively. In these two sections, we evaluate the performance of \citet{mcfadden1978modeling}'s correction factor across the entire posterior distribution irrespective of sample size, and in relation to the consistency of Bayesian point estimates for large sample sizes. Section \ref{Sec:MonteCarlo} supports our theoretical findings with the use of Monte Carlo simulations. Section \ref{Sec:Concl} concludes and sets out a pathway for future research. 
         
\section{Sampling of alternatives in classical MNL models}
\label{Sec:MNL}
\subsection{Sampling of alternatives and McFadden's correction factor}
\label{Subsec:MNL_correction}
\noindent This paper follows the conventional micro-econometric approach to individual decision-making. Individual $n$ is assumed to select alternative $i$ from the choice set $C_{n}$ when it generates the highest level of indirect utility $U_{in}$. Indirect utility in Eq.~\eqref{Eq:Util} comprises a deterministic part $V_{in}$ and an additive unobserved stochastic part $\epsilon_{in}$. The deterministic part is a function of explanatory variables $X_{in}$ and associated parameters $\beta$, and typically takes a linear form. The stochastic term $\epsilon_{in}$ is assumed to be independently and identically distributed across alternatives and individuals.  

\begin{equation}
\label{Eq:Util}
U_{in}=V_{in}+\epsilon_{in}=V(X_{in};\beta)+\epsilon_{in}
\end{equation}   

\noindent Assume that the `true' data generating process takes the form of the logit model such that $\epsilon_{in}$ follows a Type 1 Extreme Value distribution. Accordingly, $P(i|\beta,C_{n})$ in Eq.~\eqref{Eq:ChoiceProb} describes the probability that individual $n$ will select alternative $i$.\footnote{To improve the clarity of notation, we remove the conditioning on $X_{n}$ from all probability statements. $X_{n}$ typically represents a set of exogenous variables not associated with any form of stochasticity. If any such stochasticity is present, for example in a latent variable model, our Bayesian results still apply as long as the stochasticity is independent of $\beta$.}  The denominator highlights the computational challenge of applying MNL to large choice sets.

\begin{equation}
\label{Eq:ChoiceProb}
P(i|\beta, C_n)=\frac{\exp(V_{in})}{\sum_{j \in C_{n}} \exp(V_{jn})}
\end{equation}   

\noindent Sampling of alternatives reduces the computational burden by specifying a quasi-likelihood function approximating Eq.~\eqref{Eq:ChoiceProb} using the smaller choice set $D_{n}$. $D_{n}$ is a subset of $C_{n}$ and includes a number of randomly sampled alternatives, besides the chosen alternative $i$. Let $P(i|\beta,D_n)^{\dagger}$ in Eq.~\eqref{Eq:ChoiceProb1} define the \emph{uncorrected} sampled choice probability. The first aspect revealed by Eq.~\eqref{Eq:ChoiceProb1} is that by using the sampled choice set $D_{n}$, which is a subset of $C_{n}$, a smaller number of alternatives will be included in the denominator of Eq.~\eqref{Eq:ChoiceProb1} and by default an \emph{overestimate} of the true choice probability will be obtained. The second aspect revealed by Eq.~\eqref{Eq:ChoiceProb1} is that fewer utility differences are evaluated. These utility differences provide information on the parameters of interest. Accordingly, Eq.~\eqref{Eq:ChoiceProb1} may be associated with bias in its corresponding parameter estimates $\hat{\beta}$ because the sampling protocol determines the likelihood for each alternative to enter the subset $D_{n}$, and thereby the utility differences to be evaluated to learn $\beta$.   

\begin{equation}
\label{Eq:ChoiceProb1}
P(i|\beta,D_n)^{\dagger}=\frac{\exp(V_{in})}{\sum_{j \in D_{n}} \exp(V_{jn})}=\frac{1}{1+\sum_{j \neq i \in D_{n}} \exp(V_{jn}-V_{in})}
\end{equation}    

\citet{mcfadden1978modeling} proved consistent parameter estimates are obtained by adding the correction factor $ln(\pi(D_{n}|j))$ to the indirect utility function of each alternative $j \in D_{n}$, where $\pi(D_{n}|j)$ denotes the probability of sampling the choice set $D_{n}$ conditional on alternative $j$ being the chosen alternative. The only requirement for consistency is a trivial condition on this sampling probability -   which \citet{mcfadden1978modeling} labels as \emph{Positive Conditioning}, i.e. $\pi(D_{n}|j)>0 \text{ }\forall \text{ } j \in D_{n}$. A positive conditional probability of sampling the choice set $D_{n}$ allows evaluating the correction factor $ln(\pi(D_{n}|j))$. Appendix \ref{App:McFadden} replicates \citet{mcfadden1978modeling}'s original proof which, unfortunately, is not easily accessible in this digital age. 

Eq.~\eqref{Eq:McFad} presents \citet{mcfadden1978modeling}'s corrected MNL choice probability. Again, two aspects can be observed. First, in the numerator $\pi(D_{n}|i)$ scales down the choice probability and corrects for the overestimation of the choice probability. It is commonly assumed that the sampling protocol $\pi(D_{n}|i)$ is independent of $\beta$ and accordingly this effect will only rescale the likelihood (i.e. add a negative constant to the log-likelihood) of the model and does not influence where it is maximised. This effect does not correct for any potential bias in the parameters of interest. Second, in the denominator, the correction factor corrects for the potential bias induced by the sampling protocol as to which utility differences are contrasted.        

\begin{equation}
    \label{Eq:McFad}
    P(i|\beta,D_{n})=\frac{exp(V_{in}+ln(\pi(D_{n}|i)))}{\sum_{j \in D_{n}}exp(V_{jn}+ln(\pi(D_{n}|j)))}= \frac{\pi(D_{n}|i) }{\pi(D_{n}|i) + \sum_{j \neq i \in D_{n}}exp(V_{jn}-V_{in}+ln(\pi(D_{n}|j)))}
\end{equation}

\subsection{Sampling protocols}
\label{Sec:Sampling}
Besides including the chosen alternative, the analyst needs to make several practical considerations when sampling the remainder of $D_{n}$. Most notably, the sampling protocol determines the (expected) size of the sampled choice set and thereby introduces a trade-off between speed and accuracy. With smaller $D_{n}$, the estimation will be computationally more attractive, but the approximation of the denominator in the choice probability is likely to become less accurate. Notably, the number of utility differences evaluated - which provide the necessary information to estimate the parameters of interest - is reducing linearly with the size $D_{n}$. The balance between computational costs and sampling error remains an empirical matter. \citet{NerellaBhat2004} suggest that an eighth of the size of the full choice set should be used as minimum, but that a fourth is desirable.

The interested reader is referred to \citet{BenAkiva1985} for a description of various sampling protocols to obtain the target size of $D_{n}$, including the corresponding correction factors. Monte Carlo simulations conducted by \citet{Dalyetal2014} highlight that independent and with-replacement sampling allows for efficient estimation and easy calculation of $\pi(D_{n}|j) \text{ } \forall j \in D_{n}$. These two sampling protocols, however, result in variation in the size of $D_{n}$ across observations, whereas the size of $D_{n}$ is fixed under the sampling without replacement. The calculation of the correction factor $\pi(D_{n}|j)$ is, typically, more complex under the sampling without replacement and estimation is less efficient. Stratified sampling, where a fixed number of alternatives is sampled from a number of strata, suffers from the same challenges as sampling without replacement. Whereas \citet{Dalyetal2014} discard stratified sampling due to poor estimation performance, \citet{Guevara2013} apply a stratified importance sampling protocol without replacement in their Monte Carlo simulation on nested logit models in order to ensure that sufficient alternatives are sampled per nest.  

The next consideration is assigning a sampling probability to each \emph{alternative} in the choice set. It is at the level of alternatives that standard uniform or importance sampling probabilities can be assigned. In the former, sampling probabilities are constant across alternatives whereas in the latter they vary based on some specified rule. For example, in destination choice, alternatives close to the chosen alternative may receive a higher sampling probability as they are more likely to be chosen. Typically, defining the non-constant sampling probabilities at the level of alternatives is done by relying on results from previous studies such that no endogeneity is introduced. For \citet{mcfadden1978modeling}'s correction factor, the resulting conditional sampling probabilities of the \emph{set} $\pi(D_{n}|j) \text{ } \forall j\in D_{n}$ are relevant. Different combinations of sampling probabilities and protocols at the level of the alternatives may result in the same conditional sampling probabilities at the level of the set $D_{n}$. 

An attractive choice for the conditional sampling probability at the level of the set is uniform conditioning whereby an equal conditional probability of being sampled is assigned to each eligible $D_{n}$ such that $\pi(D_{n}|j)=\pi(D_{n}|k) \text{ } \forall j,k$. Uniform conditioning adds a constant to the indirect utility of each alternative in Eq.~\eqref{Eq:McFad} and thereby cancels it out in the corrected choice probability. Eq.~\eqref{Eq:ChoiceProb1} thus conveniently provides consistent estimates under uniform conditioning without the need to make adjustments to the indirect utility function. It is intuitive that no correction factor is needed under uniform conditioning to avoid bias because the sampling protocol does not steer the analysis towards a specific set of alternatives from the full choice set. Uniform conditioning does not make it more or less likely that certain utility differences are evaluated. As the sample size increases, the pure randomness of uniform conditioning increases the probability that all relevant utility differences are studied in a balanced way across the sample and consistent parameter estimates are therefore obtained.    

An undesirable feature of the uniform conditioning, however, is that in decision problems with a really large number of alternatives (e.g. destination choice), the inclusion of irrelevant options in the sampled choice set may require a prohibitively large sample size to achieve proper estimation. As noted above, importance sampling overcomes this issue by assigning a higher sampling probability to alternatives that are a priori more likely to be chosen. Importance sampling at the level of the alternative often results in a non-uniform conditional sampling probability such that $\pi(D_{n}|j)\neq \pi(D_{n}|k) \text{ } \forall j,k$. In this context, \citet{mcfadden1978modeling}'s correction factor is, however, needed as per Eq.~\eqref{Eq:McFad} to correct for the potential bias induced by steering the analysis towards a specific set of alternatives from the full choice set. In order to successfully evaluate $ln(\pi(D_{n}|j)) \text{ } \forall j \in D_{n}$ \citet{mcfadden1978modeling} has defined the positive conditioning requirement. Uniform conditioning is considered a specific case of the more generic positive conditioning.    

\subsection{Information divergence}
\label{SubSec:InfDiv}
The difference between the quasi log-likelihood based on Eq.~\eqref{Eq:McFad} and the correct log-likelihood evaluating the full choice set is studied in more detail by \citet{keane2012estimation}. Similar to \citet{mcfadden1978modeling}'s proof, \citet{keane2012estimation} study this information divergence from an \emph{ex ante} perspective  -- taking expectations over all possible chosen alternatives and all possible sampled choice sets. Appendix \ref{App:Keane} analytically derives their two main conclusions for the case of uniform conditioning: i) the expected information divergence is positive, and ii) the expected information divergence is minimised at the true parameters. In Appendix \ref{App:Keane}, we furthermore show that this result does not generalise to the case of positive conditioning. Instead, we arrive at similar but slightly adjusted conclusions in relation to \citet{mcfadden1978modeling}'s correction factor under positive conditioning: 1) the expected loss in information with respect to the parameters of interest is positive, and 2) the expected loss in information with respect to the parameters of interest is minimised at the true parameters.

The explanation for the adjusted conclusions can be traced back to the differences in the numerator and the denominator between the sampled and true choice probability. Together these two aspects form the expected information divergence between the sampled and the true log-likelihood under positive conditioning.  With reference to Section \ref{Subsec:MNL_correction}, the discrepancy in the numerator only rescales the choice probability independently of $\beta$, whereas the discrepancy in the denominator comprises the loss of information with respect to the parameters of interest. Under uniform conditioning, there is only a discrepancy in the denominator of the MNL model, not in the numerator and therefore the expected information divergence is equivalent to the expected information loss with respect to the parameters of interest. Under uniform and positive conditioning, the size of the denominator is smaller in the sampled model because fewer alternatives are included and because $ln(\pi(D_{n}|j))<0 \text{ } \forall j \in D_{n}$. This second source of discrepancy results in a positive contribution to the expected information divergence as it inflates the choice probabilities in the sampled model. For uniform conditioning, the sign of the overall expected information divergence is therefore guaranteed to be positive. For positive conditioning, the appearance of $ln(\pi(D_{n}|j))$ in the numerator, however, reduces the expected information divergence since the choice probability is scaled down. An over-correction for the overestimation of the MNL choice probability may in theory occur under positive conditioning causing the sampled likelihood to be lower than the true likelihood. A positive sign of the expected information divergence can therefore not be guaranteed for positive conditioning and \citet{keane2012estimation}'s first result therefore does not transfer to the more generic case of positive conditioning. 

Nevertheless, we show in Appendix \ref{App:Keane} that at the true parameters, the second (positive) source of expected information divergence is minimised under \emph{both} uniform and positive conditioning. Hence, \citet{mcfadden1978modeling}'s correction factor not only results in consistent parameter estimates under uniform and positive conditioning in MNL but also minimises the expected loss of information with respect to the parameters of interest. Since the first source of information divergence is independent of $\beta$, the fact that the overall expected information divergence is not necessarily minimised when using \citet{mcfadden1978modeling}'s correction factor under positive conditioning is irrelevant for estimation purposes. 

\section{Sampling of alternatives in Bayesian MNL models}          
\label{Sec:BayesMNL}
\noindent When moving from classical to Bayesian estimation, the original objective - recovering the true point parameter estimates $\beta^{*}$ - changes to recovering the posterior distribution. The posterior distribution $p(\beta|Y,C)$ in Eq.~\eqref{Eq:BayesRule_MNL} is described as a function of the prior distribution $p(\beta)$, the likelihood $p(Y|\beta,C)$ and the marginal likelihood $p(Y|C)$ - where $C$ denotes the set of full choice sets $C_{n}$ across all observations $n=1,2,\ldots,N$ and $Y$ the vector of all observed choices in the sample. 

\begin{equation}
\label{Eq:BayesRule_MNL}
p(\beta|Y,C)=\frac{p(\beta)p(Y|\beta,C)}{p(Y|C)}=\frac{p(\beta)\prod_{n=1}^{N}\frac{\exp(V_{in})}{\sum_{j\in C_{n}}\exp(V_{jn})}}{\int_{\beta}p(\beta)\prod_{n=1}^{N}\frac{\exp(V_{in})}{\sum_{j\in C_{n}}\exp(V_{jn})}d\beta}
\end{equation} 

\begin{itemize}
    \item The prior distribution $p(\beta)$ summarises all information about the parameters of interest \emph{before} collecting the data. Given our interest in using a quasi-likelihood function to approximate the true likelihood function, containing the \emph{same} $\beta$ parameters, there is no reason to assume that the prior differs between the `true' and `sampled' model.
    \item The likelihood $p(Y|\beta,C)$ is \emph{identical} to the likelihood function in classical estimation and describes the probability of observing the choices $Y$ in the sample for a given $\beta$ and full choice set $C$. The same applies to the sampled likelihood function introduced below.
    \item The marginal likelihood $p(Y|C)$ represents the expected likelihood of the model and thereby describes how well the model fits the data across all possible values of $\beta$. 
\end{itemize} 

\noindent Since $\beta$ is integrated out of the marginal likelihood, it does not contain information on the parameters of interest and only scales the posterior distribution to ensure that the density integrates to one. For estimation purposes, it is therefore often considered that the posterior is proportional to the prior and the likelihood, i.e. $p(\beta|Y,C)\propto p(\beta) \cdot p(\beta|Y,C)$. In a similar vein, any terms that are multiplicatively unrelated to $\beta$, and thereby only scale the posterior, can be removed since they do not contain information on the parameters of interest.  If we now let Eq.~\eqref{Eq:BayesRule_MNL_sampled_MCF} describe the posterior density under \citet{mcfadden1978modeling}'s correction factor, Eqs.~\eqref{Eq:BayesRule_MNL_sampled_MCF2}-\eqref{Eq:BayesRule_MNL_sampled_MCF3} simplify the posterior to only the elements which contain information with respect to the parameters of interest. It becomes directly clear that \citet{mcfadden1978modeling}'s correction factor role in the numerator is only to scale the likelihood function but is irrelevant for estimation purposes. It only corrects for the loss of information with respect to the parameters of interest through the denominator of the MNL choice probability.  

\begin{eqnarray}
\label{Eq:BayesRule_MNL_sampled_MCF}
p(\beta|Y,D)&=&\frac{p(\beta) \cdot p(Y|\beta,D)}{p(Y|D)}=\frac{p(\beta) \cdot \prod_{n=1}^{N}\frac{\exp(V_{in} + ln(\pi(D_{n}|i)))}{\sum_{j\in D_{n}}\exp(V_{jn}+ln(\pi(D_{n}|j)))}}{\int_{\beta}p(\beta) \cdot \prod_{n=1}^{N} \frac{\exp(V_{in} + ln(\pi(D_{n}|i)))}{\sum_{j\in D_{n}}\exp(V_{jn}+ln(\pi(D_{n}|j)))}d\beta}\\
p(\beta|Y,D)&\propto&p(\beta)\cdot p(Y|\beta,D)=p(\beta) \cdot \prod_{n=1}^{N}\frac{\exp(V_{in} + ln(\pi(D_{n}|i)))}{\sum_{j\in D_{n}}\exp(V_{jn}+ln(\pi(D_{n}|j)))} \label{Eq:BayesRule_MNL_sampled_MCF2}\\
p(\beta|Y,D)&\propto&p(\beta)\cdot p(Y|\beta,D)=p(\beta) \cdot \prod_{n=1}^{N}\frac{\exp(V_{in})}{\sum_{j\in D_{n}}\exp(V_{jn}+ln(\pi(D_{n}|j)))} \label{Eq:BayesRule_MNL_sampled_MCF3}
\end{eqnarray}

\subsection{Information divergence in the MNL posterior}
\label{Sec:KLMNL}
\noindent The Kullback-Leibler divergence criterion \citep{Kullback1951} is frequently used in Bayesian statistics as a measure of information loss when approximating the `true' distribution with an alternative distribution. Under the assumption of identical prior densities, the Kullback-Leibler divergence criterion ($D_{KL}$) reduces to the difference in the expected log-likelihood between the two models plus the log of the ratio of the marginal likelihoods (see Eqs.~\eqref{Eq:KLBASIC1}-\eqref{Eq:KLBASIC2}).\footnote{The $D_{KL}$ measure takes expectations with respect to $\beta$, but is conditional on observed choices and sampled choice sets.} The latter term is also known as the (inverse) \emph{Bayes Factor}, which is independent of $\beta$ and can accordingly be placed outside of the integral. The loss in information across the posterior can thus be separated into two elements, i) the loss in information with respect to the parameters of interest $\beta$; and ii) the difference in fit between the two models. Ideally, both types of information loss are minimised when using the sampled model, but for the purposes of estimation minimising only the first term in Eq.~\eqref{Eq:KLBASIC2} is essential.

\begin{eqnarray}
\label{Eq:KLBASIC1}
D_{KL}&=&\int_{\beta}p(\beta|Y,C)ln\left(\frac{p(\beta|Y,C)}{p(\beta|Y,D)}\right)d\beta=\int_{\beta}p(\beta|Y,C)ln\left(\frac{\frac{p(\beta)p(Y|\beta,C)}{p(Y|C)}}{\frac{p(\beta)p(Y|\beta,D)}{p(Y|D)}}\right)d\beta\\
\label{Eq:KLBASIC2}
D_{KL}&=&\int_{\beta}p(\beta|Y,C) ln\left(\frac{p(Y|\beta,C)}{p(Y|\beta,D)}\right)d\beta+ln\left(\frac{p(Y|D)}{p(Y|C)}\right) 
\end{eqnarray}

For estimation purposes, the term of interest is $ln\left(\frac{p(Y|\beta,C)}{p(Y|\beta,D)}\right)$, which is the negative of the information divergence studied by \citet{keane2012estimation} and covered in Appendix \ref{App:Keane} and Section \ref{SubSec:InfDiv}. Following Appendix \ref{App:Keane}, $ln\left(\frac{p(Y|\beta,C)}{p(Y|\beta,D)}\right)$ can be rewritten in Eqs.~\eqref{Eq:InfDiv_Bayes1}-\eqref{Eq:InfDiv_Bayes2} such that two familiar terms emerge again. For completeness, $\pi(D_{n}|\beta)=\sum_{j \in D_{n}} \pi(D_{n}|j) P(j|\beta,C_{n})$ is the unconditional sampling probability of the smaller choice set $D_{n} \in C_{n}$.   

\begin{eqnarray}
    ln\left(\frac{p(Y|\beta,C)}{p(Y|\beta,D)}\right)&=& \sum_{n=1}^{N} ln\left(\frac{exp(V_{in})}{\sum_{j \in C_{n}}exp(V_{jn})}\right)-ln\left(\frac{exp(V_{in}+ln(\pi(D_{n}|i))}{\sum_{j \in D_{n}}exp(V_{jn}+ln(\pi(D_{n}|j)))}\right) \label{Eq:InfDiv_Bayes1}\\
    &=&\sum_{n=1}^{N} ln\left(\frac{\sum_{j \in D_{n}}exp(V_{jn}+ln(\pi(D_{n}|j)))}{\sum_{j \in C_{n}}exp(V_{jn})}\right)-\sum_{n=1}^{N}ln\left(\pi(D_{n}|i)\right)\\
    &=&\sum_{n=1}^{N} ln\left(\sum_{j \in D_{n}} P(j|\beta,C_{n}) \cdot \pi(D_{n}|j)\right)-\sum_{n=1}^{N}ln\left(\pi(D_{n}|i)\right)\\
    &=&\sum_{n=1}^{N} ln\left(\sum_{j \in C_{n}} P(j|\beta,C_{n}) \cdot \pi(D_{n}|j)\right)-\sum_{n=1}^{N}ln\left(\pi(D_{n}|i)\right)\\
    &=&\sum_{n=1}^{N} ln\left(\pi(D_{n}|\beta)\right)-\sum_{n=1}^{N}ln\left(\pi(D_{n}|i)\right) \label{Eq:InfDiv_Bayes2}
\end{eqnarray}

In Eq.~\eqref{Eq:InfDiv_Bayes2}, the term $\sum_{n=1}^{N} ln(\pi(D_{n}|i))$ can be traced back to the difference between the numerator of the corrected and true choice probability. As argued in Section \ref{Sec:MNL} and illustrated by Eq.~\eqref{Eq:BayesRule_MNL_sampled_MCF3}, this term can be disregarded here because the term is independent of $\beta$. The only term which describes the (negative) information loss with respect to $\beta$ is $\sum_{n=1}^{N} ln\left(\pi(D_{n}|\beta)\right)$ which can be traced back to the difference between the denominator of the corrected and true choice probability. Following the same logic as applied in Appendix \ref{App:Keane}, we now aim to minimise the expected information loss with respect to $\beta$ across the entire posterior distribution (not just the point estimate). This corresponds to maximising $\mathbb{E}\left(\sum_{n=1}^{N} ln\left(\pi(D_{n}|\beta)\right)\right)$ since the sign of $\sum_{n=1}^{N} ln\left(\pi(D_{n}|\beta)\right)$ is always negative. Expectations are taken over all possible choices made, choice sets sampled, and all possible values of $\beta$. To this end, we define the joint probability of observing the vector of parameters $\beta$, the choice for alternative $i$ and sampled choice set $D_{n}$ by:\footnote{We still assume that the sampling protocol is independent of $\beta$ and without loss of generality that other explanatory variables $X_{n}$, and $C_{n}$ are assumed to be non-stochastic. Moreover, we assume that choices and sampling probabilities are independent across choice observations.}   

\begin{equation}
    \pi(\beta,i,D_{n})=p(\beta) \cdot \pi(D_{n}|\beta) \cdot P(i|\beta,D_{n}) \label{Eq:jointProb} 
\end{equation}

$\mathbb{E}\left(\sum_{n=1}^{N} ln\left(\pi(D_{n}|\beta)\right)\right)$ is then given by:

\begin{eqnarray}
    \mathbb{E}\left(\sum_{n=1}^{N} ln\left(\pi(D_{n}|\beta)\right)\right) &=& \int_{\beta}p(\beta)\sum_{n=1}^{N}\sum_{D_{n}\in C_{n}}\sum_{i \in D_{n}} \pi(D_{n}|\beta) \cdot P(i|\beta,D_{n}) \cdot ln\left(\pi(D_{n}|\beta)\right) d\beta\\
    &=& \int_{\beta}p(\beta)\sum_{n=1}^{N} \sum_{D_{n} \in C_{n}} \pi(D_n|\beta) ln(\pi(D_{n}|\beta)) d\beta
\end{eqnarray}

Since entropy $\sum_{D_{n} \in C_{n}} \pi(D_n|\beta) ln(\pi(D_{n}|\beta))$ is maximised when applying \citet{mcfadden1978modeling}'s correction term $ln(\pi(D_{n}|i))$, it minimises the expected information loss in the posterior at \emph{every} value of $\beta$ for a given sampling protocol. Therefore, the emerging posterior distribution using \citet{mcfadden1978modeling}'s correction factor -either based on uniform or positive conditioning - provides the best approximation of the shape and location of the true posterior density in Bayesian MNL models.  

\subsection{Bayesian posterior densities and point estimates}
\label{Sec:Consistent}
The above results highlight that \citet{mcfadden1978modeling}'s correction factor has good \emph{ex-ante} small and large sample properties because it minimises the expected information loss with respect to the parameters of interest under the sampling of alternatives irrespective of sample size. As the sample size increases it can be shown that the Bayesian point estimate converges to the (consistent) point estimate obtained using classical maximum likelihood estimation.

 \citet[][chapter 12]{Train2009} highlights using the Bernstein-von Mises Theorem that the posterior mean is generally considered to be the best point estimate from the posterior density under many different loss functions. He furthermore illustrates that as the sample size increases i) the posterior density converges to a Gaussian density, and ii) the posterior variance becomes the same as that of the maximum likelihood estimate because the information in the data outweighs the information contained in the prior. Additionally, the posterior mean asymptotically converges to the maximum likelihood estimate such that asymptotically the classical sampling distribution and the Bayesian posterior distribution are the same.  As a result, for all sampling protocols satisfying either a uniform or positive conditioning, \citet{mcfadden1978modeling}'s correction factor will provide consistent parameter estimates in both the classical and Bayesian settings. 
 
 As the sample size decreases two effects occur. First, posterior and classical standard errors on the parameter estimates increase because there is less information in the data regarding $\beta$ by default. Second, sampling of alternatives is likely to generate some additional degree of error because we sample fewer chosen alternatives (from the population) and choice sets $D_{n}$ (when applying sampling of alternatives) across the sample. Only when the sample size becomes sufficiently large, the actual information loss with respect to $\beta$ (in the parameter estimate and in the posterior) will converge in probability to the expected information loss (as per \citet{mcfadden1978modeling}'s proof in Appendix \ref{App:McFadden}). 
 
 Indeed, one can correct for the actual loss in information in estimation due to sampling of alternatives, but this would require evaluating the full denominator of the MNL choice probability and defies the purpose of applying sampling of alternatives. \citet{mcfadden1978modeling}'s correction protocol thus provides the best choice of correction factor without knowing which alternatives will be sampled a priori and irrespective of sample size. Determining whether sample sizes are sufficiently large, however, remains an empirical question and is no different for Bayesian and classical estimation. Similarly, determining the severity of the loss in information for the parameters of interest should be addressed empirically, including the search for optimal sampling rates \citep{Dalyetal2014}.   

\section{Sampling of alternatives in Bayesian MMNL}
\label{Sec:BayesMMNL}

\noindent We now switch our attention to the mixed multinomial logit (MMNL) model. MMNL assumes that the individual-level parameters $\beta_{n}$ are distributed over the population of interest. The mixing density $f(\beta_{n}|\theta)$ describes the distribution of preferences, where $\theta$ is the set of hyper-parameters characterising the mixing density. For example, when $\beta_{n}$ is assumed to follow a normal density then $\theta$ will include the parameters for the mean and covariance matrix. 

\subsection{Bayesian estimation of MMNL models and the role of data augmentation}
\label{Sec:MMNL_GS}
Let Eq.~\eqref{Eq:MMNL_L} describe the likelihood function of the panel MMNL model, where $\beta_{n}$ is assumed to be constant across observations by the same individual $t=1,\ldots,T$. The model reduces to the cross-sectional MMNL model for $T=1$. Estimating the parameters of interest $\theta$ can be done using Bayesian and classical estimation methods. Both estimation methods are extensively covered by \citet[][see for example chapters 6, 10 and 12]{Train2009}. Classical estimation typically approximates the (log of the) integral in Eq.~\eqref{Eq:MMNL_L} by taking a large number of draws from $f(\beta_{n}|\theta)$ and is accordingly referred to as a maximum simulated likelihood (MSL) approach. Bayesian estimation does not optimise the (same) likelihood function but requires simulation methods to characterise the posterior density for $\theta$. Section 12.6 in \citet{Train2009} provides an accessible introduction to the process of sequentially drawing from conditional posterior densities for the MMNL model - also known as Gibbs Sampling. Once the Gibbs Sampler (GS) has converged, the draws for the hyper-parameters $\theta$ describe the posterior distribution of interest. 

\begin{equation}
    \label{Eq:MMNL_L}
    P(Y|\theta,C)=\prod_{n=1}^{N}\int \prod_{t=1}^{T}P(Y_{nt}|\beta_{n},C_{nt}) f(\beta_{n}|\theta)d\beta_{n}
\end{equation}

\noindent Assuming for illustration purposes only that the MMNL model only contains normally distributed random parameters with a mean $\mu$ and covariance matrix $\Sigma$, the GS comprises three steps repeated a large number of times and starting from some arbitrary starting values for $\beta_{n} \forall n$ and $\Sigma$ \citep[see][pp 301-302]{Train2009}.\footnote{The GS can easily be extended to include fixed parameters.} 
\begin{enumerate}
    \item $\mu|\beta_{n}\forall n,\Sigma$,  conditional on values for $\Sigma$ and $\beta_{n}\forall n$, update $\mu$ by taking a draw from $p(\mu|\beta_{n},\Sigma)=f(\beta_{n}|\mu,\Sigma)p(\mu)$. Here $p(\mu|\beta_{n},\Sigma)$ describes the conditional posterior for $\mu$, the mixing density $f(\beta_{n}|\mu,\Sigma)$ is the 'likelihood' and $p(\mu)$ is the prior density on $\mu$. \citet{Train2009} shows that a multivariate normal mixing density $f(\beta_{n}|\mu,\Sigma)$ together with a multivariate normal prior density results in a normal posterior from which it is easy to take a draw.
    \item $\Sigma|\beta_{n}\forall n,\mu$, using the new value for $\mu$ and conditional on $\beta_{n}\forall n$, update $\Sigma$ by taking a draw from $p(\Sigma|\beta_{n},\mu)=f(\beta_{n}|\mu,\Sigma)p(\Sigma)$. Here $p(\Sigma|\beta_{n},\mu)$ describes the conditional posterior for $\Sigma$, the mixing density $f(\beta_{n}|\mu,\Sigma)$ is the 'likelihood' and $p(\Sigma)$ is the prior density on the covariance matrix. \citet{Train2009} shows that a multivariate normal mixing density  $f(\beta_{n}|\mu,\Sigma)$ together with an inverted Wishart normal prior density results in an inverted Wishart posterior from which it is easy to take draws. \citet{AKINC2018} discuss alternative priors that can be used without changing the three-step nature of the GS. 
    \item $\beta_{n}\forall n|\mu,\Sigma$, using the new values for $\mu$ and $\Sigma$, update each individual level parameter $\beta_{n} \forall n$ by taking a draw from $p(\beta_{n}|Y_{n},C_{n},\mu,\Sigma)=\prod_{t=1}^{T}P(Y_{nt}|\beta_{n},C_{nt})f(\beta_{n}|\mu,\Sigma)$. Here $p(\beta_{n}|Y_{n},C_{n},\mu,\Sigma)$ describes the conditional posterior for the individual level parameter for individual $n$, $P(Y_{nt}|\beta_{n},C_{nt})$ is the MNL probability of observing the choice made by individual $n$ in choice task $t$, and $f(\beta_{n}|\mu,\Sigma)$ is the prior density of $\beta_{n}$. Given the presence of the MNL probability, it is generally impossible to find a prior (i.e. mixing density) that will result in a convenient shape for the posterior distribution for which it is easy to draw from. \citet[][pp 302]{Train2009} describes how the Metropolis-Hastings algorithm can be used in this case to take suitable draws from the conditional posterior density.     
\end{enumerate}

Changes in the shape of the mixing density $f(\beta_{n}|\theta)$ do not alter the structure of the GS. It only affects the way in which the hyper-parameters are updated in Steps 1 and 2 and some of the calculations in the Metropolis-Hastings algorithm \cite[see for example][]{DeBlasi2010}. The normal density used here is therefore not a special case, and the discussion below holds without loss of generality. 

The three-step procedure in the GS explains the terminology of `hierarchical Bayes' often found in the Bayesian MMNL literature. That is, the mixing density acts as a prior on the individual-level parameter $\beta_{n}$ in Step 3, whereas in Steps 1 and 2 the mixing density acts as the likelihood and an additional layer of prior densities is required on the hyper-parameters of the mixing density. This hierarchy emerges because the individual-level parameters are not integrated out - as happens in MSL - but they are actually estimated. Step 3 takes draws for $\beta_{n}$ for each individual and at the end of the GS, the posterior for each individual-level parameter can be characterised with the stored draws. This process of estimating the latent parameters $\beta_{n}$ is also known as data augmentation \citep{Tanner1987}. The computational benefit of augmenting $\beta_{n}$ is that conditional on $\beta_{n}$, the choice probability is MNL which is easy to evaluate whilst avoiding the need for integration.

By implementing data augmentation, Bayesian MMNL models directly estimate the individual-level parameters alongside the hyperparameters of the mixing density resulting in the joint posteriors $p(\theta,\mathbf{\beta}^{+}|Y,C)$ and $p(\theta,\mathbf{\beta}^{+}|Y,D)$, where $\mathbf{\beta}^{+}$ comprises all individual-level parameters $\beta_{n}$ across all individuals. The '+' is added to avoid confusion with $\beta$ in the MNL model. \citet[][Chapter 14 pp. 239]{Chan2019} highlight that the draws from this joint posterior can be used to characterise the marginal posteriors $p(\theta|Y,C)$ and $p(\theta|Y,D)$. We will make extensive use of the latter property in deriving our results for the sampling of alternatives in Bayesian MMNL models. Our primary interest is in estimating $\theta$, and not $\beta_{n}$, which saves a lot of computer memory by not storing the draws for $\beta_{n}$.     

\subsection{Bayesian estimation of MMNL models under sampling of alternatives}
Returning back to the challenge of estimating choice models with large choice sets, note that the MNL choice probability is only part of Step 3 of the GS. Once a draw for $\beta_{n} \forall n$ is taken Steps 1 and 2 are not influenced by the choice set size. Hence, sampling of alternatives may be able to address the computational challenge arising in Step 3 of the GS by approximating the conditional posterior density $p(\beta_{n}|Y_{n},C_{n},\theta)$ by $p(\beta_{n}|Y_{n},D_{n},\theta)$. 

All that needs to be recognised at this point is that the conditional posterior density $p(\beta_{n}|Y_{n},C_{n},\theta)$ has the same structure as the posterior derived for the MNL model but at the individual level. Since our results for MNL in Section \ref{Sec:BayesMNL} apply to any sample size, \citet{mcfadden1978modeling}'s correction factor will minimise the expected loss in information in the conditional posterior for the individual-level parameter $\beta_{n}$ assuming that i) the sampled choice set $D_{nt}$ includes the chosen alternative, ii) the sampling protocol satisfies positive conditioning. No additional information loss occurs in relation to the conditional posteriors for $\theta$, i.e. steps 1 and 2 of the GS. Since the draws from the GS converge to the draws of the joint posterior (and can be used to characterise the marginal posterior density for $\theta$), \citet{mcfadden1978modeling}'s correction factor also minimises the expected information loss of the overall MMNL model.  

Appendix \ref{App:EDKL} provides a formal proof that the expected loss of information with respect to the parameters of interest in the MMNL model under data augmentation ($\beta^{*}$ and $\theta$) are minimised when using \citet{mcfadden1978modeling}'s correction factor for all sampling protocols satisfying either uniform or positive conditioning. This result is particularly encouraging as it enables researchers to combine the computational benefits of the sampling of alternatives with those of emerging computationally efficient Bayesian estimators. Namely, data augmentation - which is crucial in extending our results from MNL to MMNL - is universally applicable to other computationally efficient Bayesian estimators, such as variational Bayes \citep{BANSAL2020,rodrigues2020scaling}.   

\subsection{Contrasting sampling of alternatives for MMNL with Bayesian and MSL methods}
Bayesian posterior analysis does not rely on the asymptotic sampling distribution, the results for MNL and MMNL apply to samples of any size $N$. \citet{mcfadden1978modeling}'s correction factor will minimise the expected information loss in the full posterior density with respect to the parameters of interest, whether that is $\beta$ in MNL or $\theta$ in MMNL. This result applies to all sampling protocols satisfying either uniform or positive conditioning.  

In relation to large sample sizes, we can again invoke the Bernstein-von Mises Theorem such that as the number of respondents $N$ becomes sufficiently large the marginal posterior densities $p(\theta|Y,C)$ and $p(\theta|Y,D)$ converge to the asymptotic sampling distributions of their maximum likelihood counterparts.\footnote{\citet{DeBlasi2010} proof the consistency of Bayesian MMNL models, based on the described GS structure, and their results apply to a wide variety of mixing densities, including non-parametric ones.} MSL additionally requires the number of draws $R$ to approximate the integral in the MMNL likelihood function to rise faster than $\sqrt{N}$ in order for classical estimation of the MMNL model to be "\emph{consistent, asymptotically normal, efficient and equivalent to maximum likelihood}" \cite[][Chapter 10, pp. 256]{Train2009}. Since Bayesian estimation of the MMNL model using data augmentation does not approximate the referred integral the requirement on $R$ does not transfer to the Bayesian setting.  

Data augmentation also circumvents the challenges described by \citet{guevara2013sampling} and \citet{keane2012estimation} in relation to implementing sampling of alternatives in MMNL (and latent class) models. Following \citet{guevara2013sampling}, if we would know $\beta_{n}$ - which is the case under data augmentation - then the joint probability of observing the choice for alternative $i$ and sampled choice set $D_{nt}$ would be described by:

\begin{equation}
    \pi(D_{nt},i|\beta_{n})=\pi(D_{nt}|\beta_{n}) \cdot P(i|D_{nt},\beta_{n})
\end{equation}

Integrating out $\beta_{n}$, as required for MSL, results in the following expression conditional on the true hyper-parameters $\theta^{*}$ of the mixing density:

\begin{equation}
    \pi(D_{nt},i|\theta^{*})=\int_{\beta_{n}}\pi(D_{nt}|\beta_{n}) \cdot P(i|D_{nt},\beta_{n})f(\beta_{n}|\theta^{*})d\beta_{n}
\end{equation}

Following Bayes' rule we can accordingly define:

\begin{eqnarray}
    \label{Eq:GBA_corrected}
    \pi(i|D_{nt},\theta^{*})&=&\frac{\pi(D_{nt},i|\theta^{*})}{\pi(D_{nt}|\theta^{*})}\\
    &=&\frac{\int_{\beta_{n}}\pi(D_{nt}|\beta_{n})P(i|\beta_{n},D_{nt})f(\beta_{n}|\theta^{*})d\beta_{n}}{\pi(D_{nt}|\theta^{*})} \\
    &=&\int_{\beta_{n}}W_{nt}P(i|\beta_{n},D_{nt})f(\beta_{n}|\theta^{*})d\beta_{n}
\end{eqnarray}

where:
\begin{equation}
    W_{nt}=\frac{\pi(D_{nt}|\beta_{n})}{\pi(D_{nt}|\theta^{*})}=\frac{\sum_{j \in D_{nt}} \pi(D_{nt}|j)P(j|\beta_{n},C_{nt})}{\sum_{j \in D_{nt}} \pi(D_{nt}|j) P(j|\theta^{*},C_{nt})}
\end{equation}

\citet{guevara2013sampling} show that consistent estimates are obtained when in addition to McFadden's correction factor the term $W_{nt}$ is included in the corrected likelihood function. This corrected likelihood function is, however, not feasible because the term $W_{nt}$ still depends on the full choice set in both the numerator and the denominator. To circumvent the dependency of $W_{nt}$ on $C_{nt}$, \citet{guevara2013sampling} develop a feasible estimator. The estimator approximates $W_{nt}$ by only using elements of the sampled choice set $D_{nt}$ whilst retaining the consistency property. The necessary expansion factors can be determined by using i) population shares, ii) observed choices by the individual and iii) the naive method which sets $W_{nt}$ to 1 and thus reduces to applying only \citet{mcfadden1978modeling}'s correction factor. Since the naive approximation of $W_{nt}$ provides consistent estimates and provides good results in Monte Carlo simulations and real-world examples, this is the recommended approach by \cite{guevara2013sampling} and \cite{keane2012estimation}.

In effect, both the Bayesian and the MSL approach argue that \citet{mcfadden1978modeling} correction factor is the only necessary correction factor for applying sampling of alternatives in MMNL models. To arrive at this conclusion, both approaches take a different avenue. Bayesian estimation circumvents the problem of latent $\beta_{n}$ by augmenting the parameter and directly estimating it negating the need for an additional correction factor. Classical estimation methods, however, acknowledge the latent nature of $\beta_{n}$ and argue that in principle an additional correction factor is required, because $W_{nt} \neq 1$. In practice, the need for this additional correction factor is negligible. \citet{keane2012estimation}, for example, show in a Monte Carlo analysis that the bias introduced is very limited for modest-sized subsets $D_{nt}$. Thus, if the number of respondents becomes sufficiently large both Bayesian and MSL estimates for MMNL models using sampling of alternatives will result in consistent parameter estimates for \emph{any} sampling protocol satisfying positive conditioning when \citet{mcfadden1978modeling}'s correction factor is applied.

\section{Monte Carlo analysis}
\label{Sec:MonteCarlo}
This section presents a Monte Carlo analysis, illustrating the implementation of sampling of alternatives using Bayesian and Classical estimators of MMNL. Table \ref{Table:MCsetup} summarises the implemented simulation settings. We assume that a group of 250 or 1,000 individuals ($N$) is making a sequence of either 5 or 10 choices ($T$) each. The number of alternatives in each choice set $C_{nt}$ is 50 or 100, from which respectively 20 or 10, or 30 or 10 alternatives are sampled (including the chosen alternative). The data generating process is repeated such that for each of the 16 unique combinations of settings, 30 datasets are generated and the corresponding models are estimated. The resampling happens at the level of the choice data, not the sampling of alternatives. Both classical and Bayesian estimation of MMNL was done using the true and sampled choice sets. In the Bayesian estimation, non-informative priors are used and we take 20,000 posterior draws (burn-in: 10,000, thinning:10, effective posterior draws: L=1,000), which are sufficient for convergence as the Gelman-Rubin Diagnostic is close to 1 for all parameters. For the MSL approach, 100 draws were taken using modified Latin hypercube sampling to approximate the integral in the likelihood function \citep{Hess2006}.

Each alternative in the choice set is characterised by four attributes. Following \citet{keane2012estimation}, the first two attributes are  drawn from standard normal distributions and are associated with normally distributed random parameters with mean \{1;1\} and covariance matrix $\begin{pmatrix}
    1 & 0.6 \\
    0.6 & 1 
\end{pmatrix}$. The third and fourth attributes are dummy variables where the value of one is associated with a probability of 50\%. These two dummy variables are associated with fixed parameter values of \{1;-1\}. The sampled choice set includes the chosen alternative and the remainder of the sampled choice set is obtained using uniform sampling without replacement such that we are working in the context of uniform conditioning and no correction factor is required in practice. 

\begin{table}[!ht]
    \centering
    \caption{Monte Carlo analysis setup}
    \begin{tabular}{|l|l|l|}    
    \hline
        & Smaller choice set size & Larger choice set size \\ \hline
        Choice set size ($C_{nt}$) & 50 & 100 \\ \hline
        Size of sampled subset ($D_{nt}$) & \{20,10\} & \{30,10\} \\ \hline
        Number of individuals ($N$) & \{250,1000\} & \{250,1000\} \\ \hline
        Choice tasks per individual ($T$) & \{5,10\} & \{5,10\} \\ \hline                
        Number of MC resamples & 30 & 30 \\ \hline
        Estimation & Classical and Bayesian MMNL  & Classical and Bayesian MMNL \\ \hline
    \end{tabular}
    \label{Table:MCsetup}
\end{table}

We summarise the results of the simulation study by computing the following five metrics: (a) mean of the parameter estimates across the 30 repetitions; (b) standard deviation around this mean ; (c) average absolute percentage bias (APB); (d) average coverage probability (CP) in percentage; and (e) mean of the standard error of the parameter estimates across the 30 repetitions. Using the classical estimation setting, as an example, we compute the APB and the CP of a parameter corresponding to a synthetic dataset as follows:

\begin{eqnarray}
    APB&=& \frac{100}{R}\sum_{r=1}^{R} \left|  \frac{\hat{\beta_{r}}-\beta^{*}}{\beta^{*}}\right|  \\ 
    CP&=&\frac{100}{R}\sum_{r=1}^{R}I\left[\hat{\beta}-1.96\cdot st.error(\hat{\beta})\leq \beta^{*} \leq \hat{\beta}+1.96\cdot st.error(\hat{\beta})\right]
\end{eqnarray}

Where $I[\cdot]$ is the indicator function and R is the number of synthetic datasets which is 30 in our case. Note that the corresponding values in the Bayesian setting relate to the posterior mean and posterior standard deviation in the APB computation. To compute CP in the Bayesian estimation, we compute the percentage of synthetic datasets where the true value lies in the 95\% highest posterior density interval.  

Table \ref{tab:50-all} presents the results assuming a choice set of 50 alternatives whilst estimating standard MMNL models without a sampling of alternatives using classical and Bayesian methods. Overall, the true parameter estimates are recovered with a high level of accuracy across the 30 MC resamples. When the sample size increases either by increasing the number of respondents or the number of choices per respondent we generally see a reduction in the variability of the parameter estimates, their bias, increases in the average coverage probability and reductions in the average standard error of the parameter estimates. This confirms that for the base scenario, classical and Bayesian estimation methods have a comparable performance. Table \ref{tab:50-all} therefore acts as a point of reference in the context of sampling of alternatives applied below.    

\begin{table}[htbp]
  \centering
  \footnotesize
  \caption{True choice set size 50, no sampling of alternatives applied}
    \begin{tabular}{r r | r r | r r r r r | r r r r r }    
    \toprule
    & & & & \multicolumn{5}{c|}{Classical - full choice set} & \multicolumn{5}{c}{Bayesian - full choice set}\\
    N & T & \multicolumn{2}{c|}{True}  & Mean  & SD    & APB   & CP    & Mean & Mean  & SD    & APB   & CP    & Mean\\
     &  & \multicolumn{2}{|c|}{}  &   &     &    &     & st.er.&   &    &    &     & st.er.\\
    \midrule
 250   & 5 & B1&1.00  & 0.98  & 0.08  & 6.38\% & 93.33\% & 0.07  & 1.00  & 0.07  & 5.84\% & 93.33\% & 0.07 \\    
&&B2&1.00  & 0.97  & 0.09  & 7.50\% & 93.33\% & 0.07  & 0.99  & 0.08  & 6.33\% & 86.67\% & 0.07 \\    
&&B3&1.00  & 1.02  & 0.03  & 2.85\% & 93.33\% & 0.04  & 1.02  & 0.03  & 3.05\% & 93.33\% & 0.04 \\    
&&B4&-1.00 & -1.01 & 0.03  & 2.56\% & 90.00\% & 0.03  & -1.01 & 0.03  & 2.59\% & 93.33\% & 0.04 \\    
&&C1&0.60  & 0.60  & 0.08  & 13.86\% & 93.33\% & 0.06  & 0.57  & 0.10  & 13.72\% & 93.33\% & 0.10 \\    
&&C2&1.00  & 0.98  & 0.06  & 9.67\% & 96.67\% & 0.08  & 0.98  & 0.11  & 10.76\% & 96.67\% & 0.13 \\    
&&C3&1.00  & 0.96  & 0.07  & 11.08\% & 96.67\% & 0.06  & 0.98  & 0.13  & 8.62\% & 93.33\% & 0.13 \\    
\midrule
&10&B1&1.00  & 0.97  & 0.08  & 6.63\% & 86.67\% & 0.07  & 0.99  & 0.06  & 4.89\% & 96.67\% & 0.07 \\    
&&B2&1.00  & 0.97  & 0.07  & 6.07\% & 86.67\% & 0.07  & 0.99  & 0.06  & 4.47\% & 93.33\% & 0.07 \\    
&&B3&1.00  & 1.00  & 0.02  & 1.66\% & 90.00\% & 0.02  & 1.01  & 0.02  & 1.69\% & 93.33\% & 0.03 \\    
&&B4&-1.00 & -1.00 & 0.02  & 1.86\% & 93.33\% & 0.02  & -1.00 & 0.02  & 1.80\% & 100.00\% & 0.03 \\    
&&C1&0.60  & 0.64  & 0.06  & 12.59\% & 93.33\% & 0.05  & 0.60  & 0.08  & 16.93\% & 96.67\% & 0.09 \\    
&&C2&1.00  & 1.04  & 0.05  & 8.63\% & 86.67\% & 0.07  & 1.00  & 0.09  & 13.31\% & 100.00\% & 0.11 \\    
&&C3&1.00  & 1.03  & 0.06  & 12.08\% & 93.33\% & 0.05  & 1.01  & 0.11  & 7.19\% & 96.67\% & 0.11 \\    
\midrule
1000&5&B1&1.00  & 0.98  & 0.04  & 5.40\% & 96.67\% & 0.04  & 1.01  & 0.07  & 5.87\% & 96.67\% & 0.04 \\   
&&B2&1.00  & 0.98  & 0.04  & 5.89\% & 86.67\% & 0.04  & 1.00  & 0.07  & 5.82\% & 93.33\% & 0.04 \\    
&&B3&1.00  & 0.99  & 0.02  & 2.41\% & 90.00\% & 0.02  & 1.00  & 0.03  & 2.41\% & 96.67\% & 0.02 \\    
&&B4&-1.00 & -1.00 & 0.02  & 1.96\% & 86.67\% & 0.02  & -1.00 & 0.02  & 1.90\% & 100.00\% & 0.02 \\    
&&C1&0.60  & 0.65  & 0.05  & 12.84\% & 100.00\% & 0.03  & 0.62  & 0.08  & 9.63\% & 96.67\% & 0.05 \\    
&&C2&1.00  & 1.02  & 0.09  & 10.84\% & 93.33\% & 0.04  & 1.02  & 0.14  & 11.51\% & 93.33\% & 0.07 \\    
&&C3&1.00  & 1.05  & 0.04  & 12.08\% & 96.67\% & 0.03  & 1.05  & 0.13  & 10.51\% & 96.67\% & 0.07 \\    
\midrule
&10&B1&1.00  & 0.98  & 0.04  & 7.30\% & 86.67\% & 0.03  & 0.98  & 0.08  & 6.07\% & 93.33\% & 0.03 \\    
&&B2&1.00  & 0.98  & 0.04  & 7.57\% & 90.00\% & 0.03  & 0.99  & 0.08  & 5.98\% & 93.33\% & 0.03 \\    
&&B3&1.00  & 1.00  & 0.01  & 1.79\% & 93.33\% & 0.01  & 1.00  & 0.02  & 1.83\% & 96.67\% & 0.01 \\    
&&B4&-1.00 & -1.00 & 0.01  & 1.85\% & 96.67\% & 0.01  & -1.00 & 0.02  & 1.87\% & 93.33\% & 0.01 \\    
&&C1&0.60  & 0.62  & 0.05  & 15.89\% & 100.00\% & 0.03  & 0.61  & 0.08  & 10.61\% & 100.00\% & 0.05 \\    
&&C2&1.00  & 1.03  & 0.03  & 11.74\% & 90.00\% & 0.03  & 1.02  & 0.10  & 9.71\% & 96.67\% & 0.06 \\    
&&C3&1.00  & 1.05  & 0.03  & 11.87\% & 100.00\% & 0.02  & 1.05  & 0.11  & 8.13\% & 90.00\% & 0.06 \\
    \bottomrule
    \multicolumn{14}{l}{{\scriptsize \emph{N}: respondents; T: choice tasks; Mean: avg. estimate across 30 MC resamples; SD: St dev of estimate across 30 MC resamples}}\\
    \multicolumn{14}{l}{{\scriptsize APB: average percentage bias; CP: coverage probability; Mean st.er.: average of standard error across 30 MC resamples}}    
    \end{tabular}%
  \label{tab:50-all}%
\end{table}%

Table \ref{tab:50-20} and Table \ref{tab:50-10} present the same set of results when respectively 20 and 10 alternatives are sampled from the full set of 50 alternatives. In both cases, we observe that the true parameters can be recovered but in general, the level of precision is lower for the sampled choice sets than the full choice set. This is to be expected because of the reduced level of information about the parameters of interest in each choice task as a result of evaluating a reduced set of alternatives. Consequently, (on average) the degree of variation in the parameter estimates increases, the average percentage bias increases, average coverage probabilities decrease and the mean standard errors increase across the 30 MC resamples. This effect becomes more pronounced when sampling 10 instead of 20 alternatives. Especially in the latter case, we observe some CP values decrease to under 70\%, where good CP values are considered to be above 85\% and especially the average percentage bias is approaching 20\% for some of the parameters in the covariance matrix. These results are consistent between classical and Bayesian estimation. In this case, we would recommend that sampling 10 alternatives (20\%) is too few. One of the ways in which Bayesian estimation may circumvent such issues is when additional information is available and included through the use of informed prior densities, but we have not employed this strategy here.

\begin{table}[htbp]
  \centering
  \footnotesize
  \caption{True choice set size 50, 20 alternatives sampled}
    \begin{tabular}{r r | r r | r r r r r | r r r r r }    
    \toprule
    & & & & \multicolumn{5}{c|}{Classical - 20 sampled alts} & \multicolumn{5}{c}{Bayesian - 20 sampled alts} \\
    N & T & \multicolumn{2}{c|}{True}  & Mean  & SD    & APB   & CP    & Mean & Mean  & SD    & APB   & CP    & Mean \\
     &  & \multicolumn{2}{|c|}{}  &   &     &    &     & st.er. &   &    &    &     & st.er. \\
    \midrule
    250 & 5 & B1 & 1.00  & 1.01  & 0.08  & 6.53\% & 90.00\% & 0.07  & 1.02  & 0.08  & 6.78\% & 93.33\% & 0.08 \\
              &       & B2 & 1.00  & 1.01  & 0.09  & 6.70\% & 83.33\% & 0.08  & 1.02  & 0.09  & 6.39\% & 86.67\% & 0.08 \\
          &       & B3 & 1.00  & 1.04  & 0.04  & 4.80\% & 86.67\% & 0.04  & 1.04  & 0.04  & 5.07\% & 86.67\% & 0.04 \\
          &       & B4 & -1.00 & -1.03 & 0.04  & 3.86\% & 93.33\% & 0.04  & -1.03 & 0.04  & 4.16\% & 93.33\% & 0.04 \\
          &       & C1  & 0.60  & 0.61  & 0.08  & 14.48\% & 93.33\% & 0.07  & 0.60  & 0.10  & 13.95\% & 96.67\% & 0.11 \\
          &       & C2  & 1.00  & 0.99  & 0.07  & 10.61\% & 100.00\% & 0.09  & 1.07  & 0.13  & 9.90\% & 96.67\% & 0.15 \\
          &       & C3  & 1.00  & 1.01  & 0.08  & 12.61\% & 93.33\% & 0.07  & 1.08  & 0.17  & 13.19\% & 90.00\% & 0.15 \\  
    \midrule
    &10&B1 & 1.00  & 0.99  & 0.07  & 5.07\% & 90.00\% & 0.07  & 1.01  & 0.07  & 5.46\% & 96.67\% & 0.07 \\
&&B2 & 1.00  & 1.00  & 0.07  & 4.88\% & 96.67\% & 0.07  & 1.01  & 0.06  & 4.83\% & 90.00\% & 0.07 \\
&&B3 & 1.00  & 1.01  & 0.03  & 2.40\% & 93.33\% & 0.03  & 1.02  & 0.03  & 2.54\% & 93.33\% & 0.03 \\
&&B4 & -1.00 & -1.01 & 0.03  & 2.03\% & 93.33\% & 0.03  & -1.01 & 0.03  & 2.10\% & 96.67\% & 0.03 \\
&&C1  & 0.60  & 0.61  & 0.08  & 14.05\% & 90.00\% & 0.06  & 0.62  & 0.10  & 13.21\% & 96.67\% & 0.10 \\
&&C2  & 1.00  & 1.02  & 0.29  & 7.56\% & 90.00\% & 0.07  & 1.03  & 0.10  & 7.71\% & 93.33\% & 0.13 \\
&&C3  & 1.00  & 1.04  & 0.07  & 12.21\% & 93.33\% & 0.05  & 1.05  & 0.14  & 12.04\% & 96.67\% & 0.12 \\
    \midrule
    1000&5&B1 & 1.00  & 1.01  & 0.04  & 6.21\% & 93.33\% & 0.04  & 1.04  & 0.07  & 6.83\% & 100.00\% & 0.04 \\
&&B2 & 1.00  & 1.01  & 0.04  & 6.69\% & 93.33\% & 0.04  & 1.04  & 0.07  & 6.86\% & 96.67\% & 0.04 \\
&&B3 & 1.00  & 1.02  & 0.02  & 3.24\% & 83.33\% & 0.02  & 1.02  & 0.04  & 3.49\% & 96.67\% & 0.02 \\
&&B4 & -1.00 & -1.02 & 0.02  & 3.09\% & 86.67\% & 0.02  & -1.02 & 0.03  & 3.22\% & 93.33\% & 0.02 \\
&&C1  & 0.60  & 0.68  & 0.05  & 17.30\% & 86.67\% & 0.04  & 0.67  & 0.11  & 14.47\% & 93.33\% & 0.06 \\
&&C2  & 1.00  & 1.06  & 0.09  & 14.19\% & 90.00\% & 0.04  & 1.09  & 0.18  & 13.45\% & 86.67\% & 0.08 \\
&&C3  & 1.00  & 1.09  & 0.04  & 15.46\% & 93.33\% & 0.03  & 1.11  & 0.16  & 15.11\% & 86.67\% & 0.08 \\
\midrule
&10&B1 & 1.00  & 1.00  & 0.04  & 6.03\% & 90.00\% & 0.03  & 1.00  & 0.07  & 5.64\% & 90.00\% & 0.04 \\
&&B2 & 1.00  & 1.00  & 0.03  & 6.51\% & 96.67\% & 0.03  & 1.01  & 0.08  & 6.08\% & 93.33\% & 0.04 \\
&&B3 & 1.00  & 1.01  & 0.01  & 2.22\% & 83.33\% & 0.01  & 1.01  & 0.03  & 2.23\% & 93.33\% & 0.01 \\
&&B4 & -1.00 & -1.01 & 0.01  & 2.53\% & 90.00\% & 0.01  & -1.02 & 0.03  & 2.56\% & 90.00\% & 0.01 \\
&&C1  & 0.60  & 0.65  & 0.05  & 14.00\% & 83.33\% & 0.03  & 0.62  & 0.08  & 11.20\% & 100.00\% & 0.05 \\
&&C2  & 1.00  & 1.06  & 0.03  & 9.41\% & 80.00\% & 0.04  & 1.04  & 0.11  & 8.72\% & 93.33\% & 0.06 \\
&&C3  & 1.00  & 1.07  & 0.03  & 10.84\% & 76.67\% & 0.03  & 1.07  & 0.11  & 10.49\% & 93.33\% & 0.06 \\
    \bottomrule
    \multicolumn{14}{l}{{\scriptsize \emph{N}: respondents; T: choice tasks; Mean: avg. estimate across 30 MC resamples; SD: St dev of estimate across 30 MC resamples}}\\
    \multicolumn{14}{l}{{\scriptsize APB: average percentage bias; CP: coverage probability; Mean st.er.: average of standard error across 30 MC resamples}}        
    \end{tabular}%
  \label{tab:50-20}%
\end{table}%

Table \ref{tab:100alts} presents the results for the Bayesian analysis related to the setting with 100 alternatives. The results for classical estimation are similar and available upon request from the authors. Again, the model using the full choice set is able to recover true parameters with relatively low levels of bias and acceptable levels of the average coverage probability. When sampling 30 out of 100 alternatives, we can see the levels for APB increase and the CP fall to lower levels across the different settings. At this sampling rate, the results are close to being acceptable, but when reducing the sampling to 10 out of 100 alternatives, we can clearly see CP levels falling across the board to unacceptable levels indicating that \citet{NerellaBhat2004} suggestion of using a sampling rate of around 25\% also holds in this context when accounting for covariances across the random parameters.           
It is interesting to see that across the models presented the trend is consistent irrespective of sample size. Indeed, there is some additional bias and estimation imprecision in the context of the smaller sample sizes, but no specific trend emerges that in case of a decrease in sample sizes, higher sampling rates need to be used for sampling alternatives to be successful. We take this as supporting evidence of our theoretical result that \citet{mcfadden1978modeling}'s correction factor also has desirable small sample properties.   

\begin{table}[htbp]
  \centering
  \footnotesize
  \caption{True choice set size 50, 10 alternatives sampled}
    \begin{tabular}{r r | r r | r r r r r | r r r r r }    
    \toprule
    & & & & \multicolumn{5}{c|}{Classical - full choice set} & \multicolumn{5}{c}{Bayesian - full choice set}\\
    N & T & \multicolumn{2}{c|}{True}  & Mean  & SD    & APB   & CP    & Mean & Mean  & SD    & APB   & CP    & Mean\\
     &  & \multicolumn{2}{|c|}{}  &   &     &    &     & st.er.&   &    &    &     & st.er.\\
	 \midrule    
250 & 5&	B1    & 1.00  & 1.04  & 0.10  & 8.49\% & 86.67\% & 0.08  & 1.05  & 0.09  & 8.84\% & 86.67\% & 0.09 \\    
&&    B2    & 1.00  & 1.03  & 0.10  & 7.75\% & 90.00\% & 0.08  & 1.04  & 0.10  & 8.41\% & 90.00\% & 0.09 \\    
&&    B3    & 1.00  & 1.07  & 0.05  & 7.19\% & 73.33\% & 0.05  & 1.07  & 0.05  & 7.79\% & 66.67\% & 0.05 \\    
&&    B4    & -1.00 & -1.06 & 0.05  & 6.77\% & 70.00\% & 0.05  & -1.07 & 0.05  & 7.29\% & 70.00\% & 0.05 \\    
&&    C1    & 0.60  & 0.63  & 0.10  & 18.89\% & 90.00\% & 0.08  & 0.63  & 0.11  & 17.30\% & 93.33\% & 0.13 \\    
&&    C2    & 1.00  & 1.03  & 0.41  & 11.02\% & 96.67\% & 0.10  & 1.02  & 0.14  & 11.43\% & 96.67\% & 0.18 \\    
&&    C3    & 1.00  & 1.04  & 0.09  & 14.40\% & 86.67\% & 0.08  & 1.03  & 0.16  & 15.71\% & 100.00\% & 0.18 \\    
\midrule
&10&    B1    & 1.00  & 1.00  & 0.07  & 5.98\% & 90.00\% & 0.07  & 1.03  & 0.07  & 6.17\% & 93.33\% & 0.08 \\    
&&    B2    & 1.00  & 1.01  & 0.07  & 4.99\% & 93.33\% & 0.07  & 1.03  & 0.07  & 5.26\% & 90.00\% & 0.08 \\    
&&    B3    & 1.00  & 1.03  & 0.04  & 3.95\% & 83.33\% & 0.03  & 1.04  & 0.04  & 4.18\% & 83.33\% & 0.03 \\    
&&    B4    & -1.00 & -1.03 & 0.03  & 3.28\% & 90.00\% & 0.03  & -1.03 & 0.03  & 3.47\% & 90.00\% & 0.03 \\    
&&    C1    & 0.60  & 0.68  & 0.07  & 16.32\% & 90.00\% & 0.06  & 0.64  & 0.09  & 13.68\% & 96.67\% & 0.11 \\    
&&    C2    & 1.00  & 1.10  & 0.06  & 12.52\% & 86.67\% & 0.08  & 1.08  & 0.10  & 10.33\% & 96.67\% & 0.14 \\    
&&    C3    & 1.00  & 1.09  & 0.07  & 12.44\% & 93.33\% & 0.06  & 1.07  & 0.14  & 11.94\% & 100.00\% & 0.14 \\    
\midrule
1000&5&    B1    & 1.00  & 1.03  & 0.04  & 7.88\% & 90.00\% & 0.04  & 1.06  & 0.08  & 8.52\% & 93.33\% & 0.04 \\    
&&    B2    & 1.00  & 1.03  & 0.05  & 8.98\% & 80.00\% & 0.04  & 1.06  & 0.09  & 9.64\% & 93.33\% & 0.04 \\    
&&    B3    & 1.00  & 1.04  & 0.03  & 6.23\% & 53.33\% & 0.02  & 1.06  & 0.04  & 6.67\% & 86.67\% & 0.02 \\    
&&    B4    & -1.00 & -1.04 & 0.02  & 5.17\% & 63.33\% & 0.02  & -1.05 & 0.04  & 5.68\% & 86.67\% & 0.02 \\    
&&    C1    & 0.60  & 0.66  & 0.07  & 18.11\% & 80.00\% & 0.04  & 0.68  & 0.14  & 19.16\% & 93.33\% & 0.07 \\    
&&    C2    & 1.00  & 1.04  & 0.08  & 14.10\% & 80.00\% & 0.05  & 1.10  & 0.17  & 14.82\% & 86.67\% & 0.09 \\    
&&    C3    & 1.00  & 1.10  & 0.05  & 13.95\% & 96.67\% & 0.04  & 1.15  & 0.18  & 17.78\% & 93.33\% & 0.09 \\    
\midrule
&10&    B1    & 1.00  & 1.02  & 0.04  & 7.28\% & 93.33\% & 0.04  & 1.02  & 0.08  & 6.63\% & 90.00\% & 0.04 \\    
&&    B2    & 1.00  & 1.01  & 0.04  & 6.97\% & 93.33\% & 0.04  & 1.04  & 0.08  & 6.63\% & 93.33\% & 0.04 \\    
&&    B3    & 1.00  & 1.02  & 0.02  & 4.85\% & 63.33\% & 0.02  & 1.05  & 0.04  & 5.12\% & 66.67\% & 0.02 \\    
&&    B4    & -1.00 & -1.03 & 0.02  & 4.34\% & 63.33\% & 0.02  & -1.05 & 0.03  & 4.71\% & 80.00\% & 0.02 \\    
&&    C1    & 0.60  & 0.67  & 0.05  & 21.24\% & 76.67\% & 0.03  & 0.66  & 0.12  & 18.61\% & 93.33\% & 0.05 \\    
&&    C2    & 1.00  & 1.10  & 0.03  & 15.95\% & 86.67\% & 0.04  & 1.11  & 0.16  & 14.81\% & 80.00\% & 0.07 \\    
&&    C3    & 1.00  & 1.10  & 0.03  & 14.64\% & 90.00\% & 0.03  & 1.12  & 0.16  & 15.21\% & 83.33\% & 0.07 \\ 
    \bottomrule
    \multicolumn{14}{l}{{\scriptsize \emph{N}: respondents; T: choice tasks; Mean: avg. estimate across 30 MC resamples; SD: St dev of estimate across 30 MC resamples}}\\
    \multicolumn{14}{l}{{\scriptsize APB: average percentage bias; CP: coverage probability; Mean st.er.: average of standard error across 30 MC resamples}}    
    \end{tabular}%
  \label{tab:50-10}%
\end{table}%

\begin{landscape}
    \begin{table}[htbp]
  \centering
  \footnotesize
  \caption{True choice set size 100}
    \begin{tabular}{r r | r r | r r r r r | r r r r r | r r r r r }    
    \toprule
    & & & & \multicolumn{5}{c|}{Bayesian - full choice set} & \multicolumn{5}{c}{Bayesian - 30 alts sampled} & \multicolumn{5}{c}{Bayesian - 10 alts sampled}\\
    N & T & \multicolumn{2}{c|}{True}  & Mean  & SD    & APB   & CP    & Mean & Mean  & SD    & APB   & CP    & Mean & Mean  & SD    & APB   & CP    & Mean\\
     &  & \multicolumn{2}{|c|}{}  &   &     &    &     & st.er.&   &    &    &     & st.er.&   &    &    &     & st.er.\\
	 \midrule 
250&5&	B1 & 1.00  & 1.00  & 0.04  & 3.40\% & 100.00\% & 0.08  & 1.02  & 0.04  & 3.73\% & 90.00\% & 0.09  & 1.04  & 0.04  & 4.59\% & 86.67\% & 0.08 \\    
&&    B2 & 1.00  & 0.99  & 0.04  & 3.31\% & 96.67\% & 0.07  & 1.02  & 0.04  & 3.40\% & 93.33\% & 0.09  & 1.04  & 0.05  & 4.83\% & 83.33\% & 0.07 \\    
&&    B3 & 1.00  & 1.00  & 0.02  & 1.67\% & 90.00\% & 0.03  & 1.02  & 0.02  & 2.25\% & 80.00\% & 0.05  & 1.04  & 0.03  & 4.47\% & 53.33\% & 0.04 \\    
&&    B4 & -1.00 & -1.00 & 0.02  & 1.56\% & 96.67\% & 0.03  & -1.02 & 0.02  & 2.37\% & 83.33\% & 0.05  & -1.04 & 0.02  & 4.49\% & 60.00\% & 0.04 \\    
&&    C1  & 0.60  & 0.60  & 0.07  & 8.73\% & 83.33\% & 0.10  & 0.64  & 0.07  & 10.13\% & 90.00\% & 0.14  & 0.66  & 0.09  & 13.76\% & 83.33\% & 0.12 \\    
&&    C2  & 1.00  & 1.00  & 0.07  & 6.65\% & 93.33\% & 0.13  & 1.04  & 0.08  & 7.47\% & 83.33\% & 0.19  & 1.07  & 0.09  & 9.25\% & 83.33\% & 0.16 \\    
&&    C3  & 1.00  & 1.01  & 0.08  & 5.42\% & 93.33\% & 0.13  & 1.06  & 0.09  & 8.82\% & 90.00\% & 0.18  & 1.09  & 0.11  & 10.83\% & 90.00\% & 0.15 \\    
	\midrule	
&10&    B1 & 1.00  & 1.00  & 0.03  & 2.61\% & 100.00\% & 0.07  & 1.02  & 0.03  & 3.14\% & 96.67\% & 0.08  & 1.04  & 0.04  & 4.29\% & 86.67\% & 0.08 \\    
&&    B2 & 1.00  & 1.00  & 0.03  & 2.73\% & 96.67\% & 0.07  & 1.02  & 0.03  & 2.98\% & 93.33\% & 0.08  & 1.03  & 0.04  & 3.96\% & 86.67\% & 0.07 \\    
&&    B3 & 1.00  & 1.00  & 0.01  & 1.22\% & 90.00\% & 0.02  & 1.01  & 0.01  & 1.26\% & 90.00\% & 0.03  & 1.02  & 0.02  & 2.48\% & 66.67\% & 0.04 \\    
&&    B4 & -1.00 & -1.00 & 0.01  & 0.93\% & 93.33\% & 0.02  & -1.01 & 0.01  & 1.61\% & 86.67\% & 0.03  & -1.03 & 0.02  & 2.83\% & 60.00\% & 0.04 \\    
&&    C1  & 0.60  & 0.60  & 0.04  & 5.62\% & 93.33\% & 0.09  & 0.62  & 0.05  & 6.92\% & 90.00\% & 0.11  & 0.64  & 0.05  & 7.85\% & 90.00\% & 0.10 \\    
&&    C2  & 1.00  & 1.01  & 0.06  & 3.45\% & 90.00\% & 0.11  & 1.04  & 0.07  & 6.24\% & 100.00\% & 0.15  & 1.06  & 0.08  & 7.74\% & 90.00\% & 0.12 \\    
&&    C3  & 1.00  & 1.00  & 0.05  & 4.98\% & 96.67\% & 0.11  & 1.04  & 0.05  & 5.38\% & 90.00\% & 0.15  & 1.06  & 0.05  & 6.76\% & 76.67\% & 0.12 \\    
\midrule	
1000&5&    B1 & 1.00  & 1.00  & 0.03  & 2.44\% & 100.00\% & 0.04  & 1.02  & 0.03  & 3.03\% & 90.00\% & 0.04  & 1.05  & 0.04  & 5.05\% & 73.33\% & 0.04 \\    
&&    B2 & 1.00  & 1.00  & 0.03  & 2.68\% & 96.67\% & 0.04  & 1.02  & 0.04  & 3.09\% & 93.33\% & 0.04  & 1.05  & 0.04  & 4.82\% & 83.33\% & 0.03 \\    
&&    B3 & 1.00  & 1.00  & 0.01  & 1.04\% & 96.67\% & 0.02  & 1.02  & 0.02  & 2.28\% & 83.33\% & 0.03  & 1.07  & 0.02  & 6.53\% & 26.67\% & 0.02 \\    
&&    B4 & -1.00 & -1.00 & 0.02  & 1.45\% & 90.00\% & 0.02  & -1.02 & 0.02  & 2.14\% & 80.00\% & 0.03  & -1.05 & 0.03  & 5.41\% & 43.33\% & 0.01 \\    
&&    C1  & 0.60  & 0.59  & 0.05  & 7.56\% & 96.67\% & 0.05  & 0.62  & 0.05  & 8.21\% & 96.67\% & 0.07  & 0.65  & 0.06  & 11.07\% & 96.67\% & 0.06 \\    
&&    C2  & 1.00  & 1.01  & 0.06  & 3.93\% & 90.00\% & 0.06  & 1.05  & 0.07  & 6.90\% & 100.00\% & 0.09  & 1.09  & 0.09  & 11.29\% & 90.00\% & 0.07 \\    
&&    C3  & 1.00  & 1.00  & 0.05  & 4.81\% & 100.00\% & 0.06  & 1.04  & 0.05  & 5.67\% & 93.33\% & 0.09  & 1.08  & 0.07  & 8.59\% & 90.00\% & 0.07 \\
\midrule    
&10&    B1 & 1.00  & 1.00  & 0.04  & 3.25\% & 86.67\% & 0.03  & 1.02  & 0.04  & 3.53\% & 93.33\% & 0.04  & 1.03  & 0.04  & 4.43\% & 83.33\% & 0.04 \\    
&&    B2 & 1.00  & 1.00  & 0.03  & 2.33\% & 96.67\% & 0.03  & 1.02  & 0.03  & 2.67\% & 96.67\% & 0.04  & 1.04  & 0.03  & 4.08\% & 86.67\% & 0.03 \\    
&&    B3 & 1.00  & 1.00  & 0.01  & 0.91\% & 100.00\% & 0.01  & 1.02  & 0.01  & 1.67\% & 83.33\% & 0.02  & 1.04  & 0.01  & 3.80\% & 36.67\% & 0.02 \\    
&&    B4 & -1.00 & -1.00 & 0.01  & 0.98\% & 100.00\% & 0.01  & -1.01 & 0.01  & 1.41\% & 86.67\% & 0.02  & -1.04 & 0.02  & 3.84\% & 40.00\% & 0.01 \\    
&&    C1  & 0.60  & 0.60  & 0.05  & 7.37\% & 93.33\% & 0.04  & 0.62  & 0.06  & 8.80\% & 86.67\% & 0.06  & 0.64  & 0.07  & 10.63\% & 80.00\% & 0.05 \\    
&&    C2  & 1.00  & 1.01  & 0.07  & 4.98\% & 90.00\% & 0.06  & 1.04  & 0.08  & 6.91\% & 90.00\% & 0.07  & 1.06  & 0.08  & 8.23\% & 80.00\% & 0.06 \\    
&&    C3  & 1.00  & 1.01  & 0.06  & 5.61\% & 100.00\% & 0.06  & 1.04  & 0.06  & 5.70\% & 73.33\% & 0.07  & 1.07  & 0.07  & 7.71\% & 80.00\% & 0.06 \\
    \bottomrule
    \multicolumn{19}{l}{{\scriptsize \emph{N}: respondents; T: choice tasks; Mean: avg. estimate across 30 MC resamples; SD: St dev of estimate across 30 MC resamples}}\\
    \multicolumn{19}{l}{{\scriptsize APB: average percentage bias; CP: Cerage probability; Mean st.er.: average of standard error across 30 MC resamples}}    
    \end{tabular}%
  \label{tab:100alts}%
\end{table}%
\end{landscape}

\section{Conclusions}
\label{Sec:Concl}
In this paper, we have revisited \citet{mcfadden1978modeling}'s correction factor for the sampling of alternatives. Our analysis has gone beyond the well-known result that the correction factor results in consistent parameter estimates in the context of multinomial logit (MNL) models. The relation between the correction factor and the expected information loss with respect to the parameters of interest has been the centre of our attention. Building on the work of \citet{keane2012estimation}, we have shown that for both uniform and positive conditioning the expected loss of information - which is relevant for estimation purposes - is minimised at the true parameter values. 

We have provided an intuitive explanation of the source of the information loss with respect to the parameters of interest. Namely, since the sampling of alternatives entails a smaller number of utility differences in the denominator of the MNL choice probability, less information about the parameters of interest is obtained. Moreover, the sampling protocol may introduce bias in these parameter estimates by influencing the subset of alternatives against which the chosen alternative is most likely contrasted. \citet{mcfadden1978modeling}'s correction factor accounts for the latter effect. Because uniform conditioning does not steer the sampling to a specific set of alternatives to contrast against the utility of the chosen alternative, it makes intuitive sense that \citet{mcfadden1978modeling}'s correction factor cancels out and is not required in practice for such sampling protocols. 

We have furthermore shown that this (expected) information loss with respect to the parameters of interest is an integral part of the Kullback-Leibler divergence criterion frequently used in Bayesian statistics to measure the loss of information between the true posterior density and an alternative approximation - based on the MNL likelihood under the sampling of alternatives in our case. In fact, we argue that for estimation purposes - which aims to learn about the parameters of interest - this is the only relevant term to consider. Accordingly, we were able to establish that \citet{mcfadden1978modeling}'s correction factor minimises the expected loss in information with respect to the parameters of interest at every possible parameter value and therefore across the entire posterior density. The Bayesian MNL posterior based on \citet{mcfadden1978modeling}'s correction factor under the sampling of alternatives is therefore the best approximation of the true posterior - in terms of minimum expected information loss - irrespective of sample size. As sample sizes decrease, the amount of information in the true and sampled model reduces and the degree of uncertainty increases (i.e. increased standard errors and bias in the parameter estimates). This happens irrespective of using a Bayesian or classical maximum likelihood approach. The only way by which Bayesian models could counteract such effects is by increasing the information content of the prior, i.e. by making use of informative priors. We have furthermore established that as the sample size increases the corresponding Bayesian point estimate will be consistent. \citet{mcfadden1978modeling}'s correction factor therefore has desirable small and large sample properties. The fact that sampling of alternatives transfers to Bayesian estimation methods together with desirable finite sample performance is an important contribution to the literature.   

We continued our analysis by arguing that these convenient properties directly transfer to Bayesian MMNL models when data augmentation \citep{Tanner1987} is applied. By treating the individual level parameters as observed, the need for additional correction factors in MMNL, as discussed by \citet{guevara2013sampling}, disappears. Namely, since \citet{mcfadden1978modeling} correction factor minimises the expected loss of information with respect to the individual-level parameters and no additional information loss occurs at the level of the parameters describing the mixing density, minimum overall expected information loss is obtained irrespective of sample size. This result is particularly encouraging as it enables researchers to combine the computational benefits of the sampling of alternatives with those of emerging computationally efficient Bayesian estimators. Namely, data augmentation is universally applicable to other computationally efficient Bayesian estimators, such as variational Bayes \citep{BANSAL2020,rodrigues2020scaling}.

Notably, both the Bayesian and the MSL approach argue that \citet{mcfadden1978modeling} correction factor is the only necessary correction factor for applying sampling of alternatives in MMNL models. To arrive at this conclusion, both approaches take a different avenue. Bayesian estimation circumvents the problem of latent $\beta_{n}$ by augmenting the parameter and directly estimating it negating the need for an additional correction factor. Classical estimation methods, however, acknowledge the latent nature of $\beta_{n}$ and argue that in principle an additional correction factor is required. In practice, the need for this additional correction factor is negligible. Our contributions do not explain the good performance of this feasible Naive estimator in classical estimation \citep{Azaiez2010,keane2012estimation,Lemp2012,von2018estimation,guevara2013sampling}. If the number of respondents becomes sufficiently large, both Bayesian and MSL point estimates for MMNL models using sampling of alternatives will result in consistent parameter estimates for \emph{any} sampling protocol satisfying positive conditioning when \citet{mcfadden1978modeling}'s correction factor is applied.

We finally presented Monte Carlo analyses supporting the theoretical findings highlighted above that sampling of alternatives together with \citet{mcfadden1978modeling}'s correction factor can successfully be applied in Bayesian MMNL models. The use of alternative estimation methods, however, does not circumvent the challenges associated with finding appropriate sampling strategies, i.e. what sampling protocol to choose and how many alternatives to sample, among others. For example, uniform conditioning is desirable due to not needing to calculate the correction factor but may have undesirable properties in some empirical studies due to the inclusion of a large number of irrelevant alternatives in the sampled choice set. This can be overcome by implementing sampling protocols satisfying positive conditioning, but the calculation of the required correction factor is more complicated. Moreover, our results have shown that optimal sampling rates in a Bayesian context are likely to be comparable to the recommendations made by \citet{NerellaBhat2004} in the context of classical estimation of MMNL models. More research is needed to answer the empirical question related to the best approach, which can be addressed in future empirical and simulation-based studies. 

Our results can easily be extended to the context of a latent class model, where class memberships, instead of individual-level parameters, would be augmented. Conditional on the class membership, class-specific choice probabilities again reduce to MNL specifications, allowing for a similar exposition on minimum expected information loss. Given that \citet{von2018estimation} apply uniform conditioning in their Monte Carlo simulations and empirical work on sampling of alternatives for latent class models using the expectation-maximisation (EM) algorithm, good small sample performance can be expected and is found by the referred authors. Our contributions do not (yet) extend to the context of Multivariate Extreme Value (MEV) models \citet{Guevara2013} and Random Regret Minimisation models \citet{GuevaraChorus2015}. The additional correction factors required in these model specifications are a direct result of no longer satisfying the axiom of Independence of Irrelevant Alternatives (IIA). They cannot be circumvented by the use of data augmentation since there are no latent variables driving the need for these correction factors. Indeed, one could approximate MEV model structures with MMNL-based error components models. Alternatively, the information loss associated with and the performance of the referred correction factors can be studied in future research. 
  

\bibliographystyle{apalike}
\bibliography{bibliography.bib}

\newpage

\appendix 
\section{Consistent MNL parameter estimates under the sampling of alternatives}
\label{App:McFadden}
When introducing sampling of alternatives and its corresponding correction factor, most papers derive \citet{mcfadden1978modeling}'s result using Bayes' rule \citep[e.g.][]{BenAkiva1985,guevara2013sampling}: 

\begin{eqnarray}
\label{Eq:McFadBayes}
P(i|\beta,D_{n},X_{n})&=&\frac{\pi(D_{n}|i,X_{n})P(i|\beta,C_{n},X_{n})}{\pi(D_{n}|X_{n})}=\frac{\pi(D_{n}|i,X_{n})P(i|\beta,C_{n},X_{n})}{\sum_{j \in C_{n}} \pi(D_{n}|j,X_{n})P(j|\beta,C_{n},X_{n})}\\
&=&\frac{\pi(D_{n}|i,X_{n})P(i|\beta,C_{n},X_{n})}{\sum_{j \in D_{n}} \pi(D_{n}|j,X_{n})P(j|\beta,C_{n},X_{n})}=\frac{\pi(D_{n}|i,X_{n})\frac{\exp(V_{in})}{\sum_{k \in C_{n}}\exp(V_{kn})}}{\sum_{j \in D_{n}} \pi(D_{n}|j,X_{n})\frac{\exp(V_{jn})}{\sum_{k \in C_{n}}\exp(V_{kn})}}\\
\label{Eq:McFadBayes1}
&=&\frac{\pi(D_{n}|i,X_{n})\exp(V_{in})}{\sum_{j \in D_{n}} \pi(D_{n}|j,X_{n})\exp(V_{jn})}=\frac{\exp(V_{in}+ln(\pi(D_{n}|i,X_{n})))}{\sum_{j \in D_{n}}\exp(V_{jn}+ln(\pi(D_{n}|j,X_{n})))}	
\end{eqnarray}

In the above, $P(i|\beta,D_{n},X_{n})$ represents the corrected MNL choice probability under the sampling of alternatives, $\pi(D_{n}|i,X_{n})$ is the conditional probability of sampling the set of alternatives $D_{n}$ from the full choice set $C_{n}$, $\pi(D_{n}|X_{n})$ is the unconditional probability of sampling the set $D_{n}$ from $C_{n}$, and  $P(i|\beta,C_{n},X_{n})$ is the MNL choice probability evaluated over the full choice set. Finally, $\beta$ represents the vector of parameters of interest and $X_{n}$ is the relevant explanatory variables for observation $n$. 

\citet{mcfadden1978modeling} distinguishes two forms of probability distributions for $\pi(D_{n}|i,X_{n})$, respectively positive and uniform conditioning. Following \citet{Dalyetal2014}, positive conditioning requires the chosen alternative to be included in $D_{n}$ \emph{and} a positive conditional sampling probability $\pi(D_{n}|j,X_{n})>0 \forall j \in D_{n}$. Uniform conditioning assumes that if $i,j\in D_{n}\subset C_{n}$ then $\pi(D_{n}|i,X_{n})=\pi(D_{n}|j,X_{n})$. The equality of the conditional sampling probability under uniform conditioning causes the correction factor to cancel out such that $P(i|\beta,D_{n},X_{n})$ reduces to Eq.~\eqref{Eq:ChoiceProb1}.  

\citet{mcfadden1978modeling} proved that when the data generating process is MNL with true parameters $\beta^{*}$, i.e. when $P(i|\beta^{*},C_{n},X_{n})$ represents the true choice probability, then estimating MNL models using $P(i|\beta,D_{n},X_{n})$ to approximate the true choice probability will result in consistent parameter estimates. The original proof is close to the presentation used by \citet{keane2012estimation} in their appendix on sampling of alternatives. \citet{keane2012estimation}'s presentation, however, only covers uniform conditioning whereas \citet{mcfadden1978modeling}'s proof also applies to the more generic setting of positive conditioning and derives uniform conditioning as a special case.      

Define the corrected (or quasi) log-likelihood by:

\begin{equation}
    L_{N}=\frac{1}{N}\sum_{n=1}^{N}ln(P(i|\beta,D_{n},X_{n}))
\end{equation}

where $P(i|\beta,D_{n},X_{n})$ is as defined above. Furthermore define $\plim_{n \rightarrow \infty} L_{N}=L$ by:

\begin{equation}
    L=\int_{X_{n}}\left[\sum_{i \in C_{n}}\sum_{D_{n} \in C_{n}} P(i|\beta^{*},C_{n},X_{n})\cdot \pi(D_{n}|i,X_{n})\cdot ln\left(P(i|\beta,D_{n},X_{n})\right)\right]p(X_{n})dX_{n}    
\end{equation}

where $p(X_{n})$ is the frequency distribution of $X_{n}$.\footnote{\citet{keane2012estimation} treat $X_{n}$ as non-stochastic which does not change the outcome of the proof.} The joint density $P(i|\beta^{*},C_{n},X_{n})\cdot \pi(D_{n}|i,X_{n})$ can be rewritten using Bayes' Rule to $P(i|\beta^{*},D_{n},X_{n})\cdot \pi(D_{n}|X_{n},\beta^{*})$. Importantly, $\pi(D_{n}|X_{n},\beta^{*})$, as defined in the denominator of Eq.~\eqref{Eq:McFadBayes}-~\eqref{Eq:McFadBayes1}, depends on the true parameters not those that need to be estimated. 

\begin{eqnarray}
    L&=&\int_{X_{n}}\left[\sum_{D_{n} \in C_{n}}\sum_{i \in D_{n}} P(i|\beta^{*},D_{n},X_{n})\cdot \pi(D_{n}|X_{n},\beta^{*})\cdot ln\left(P(i|\beta,D_{n},X_{n})\right)\right]p(X_{n})dX_{n}\\
    &=& \int_{X_{n}}\left[\sum_{D_{n} \in C_{n}} \pi(D_n|X_{n},\beta^{*})
     \sum_{i \in D_{n}} P(i|\beta^{*},D_{n},X_{n})\cdot  ln\left(P(i|\beta,D_{n},X_{n})\right)\right]p(X_{n})dX_{n}   
\end{eqnarray}
        
Since $\pi(D_{n}|X_{n},\beta^{*})$ only depends on the true parameters, only $\sum_{i\in D_{n}} P(i|\beta^{*},D_{n},X_{n}) \cdot ln\left(P(i|\beta,D_{n},X_{n})\right)$ is relevant when maximising $L$ with respect to $\beta$. Since $\sum_{i \in D_{n}} P(i|\beta,D_{n},X_{n})=1$, $L$ reaches its maximum at $\beta=\beta^{*}$ because of maximum entropy. Under normal regularity conditions, this maximum is unique and it can be shown that the maxima of $L_{N}$ converge in probability to the maximum of $L$ \citep{mcfadden1978modeling}. Sampling of alternatives yields consistent estimators when using the corrected log-likelihood $L_{N}$, i.e. when applying \citet{mcfadden1978modeling}'s correction factor under uniform and positive conditioning. 

\section{Information divergence under the sampling of alternatives}
\label{App:Keane}
\citet{keane2012estimation} examine the expected difference, or information divergence, between the quasi log-likelihood and the 'true' log-likelihood. For uniform conditioning, \citet{keane2012estimation} state that the expected (positive) information divergence is minimised at $\beta=\beta^{*}$ since the difference between the quasi and 'true' log-likelihood only shifts up the expected log-likelihood, but does not alter where it is maximised. In what follows, we study the expected information divergence from the more general perspective of positive conditioning and highlight that the results from \citet{keane2012estimation} are not directly transferable. Where uniform conditioning is guaranteed to minimise the full expected information divergence, positive conditioning is only guaranteed to minimise the expected information loss with respect to the parameters of interest. 

Let Eq.~\eqref{Eq:Divergence}-\eqref{Eq:Divergence_end} define the expected information divergence between the quasi log-likelihood $(LL^{+})$ under positive conditioning and the 'true' log-likelihood $(LL)$. For notational convenience we drop the conditionality on $X_{n}$.

\begin{eqnarray}
\mathbb{E}\left(LL^{+}-LL\right) 
&=& \sum_{n}\sum_{D_{n}\in C_{n}} \sum_{i\in D_{n}} \pi(D_{n}|\beta^{*}) \cdot P\left(i|\beta^{*},D_{n}\right)\cdot  \left[ln\left(P(i|\beta,D_{n})\right) - ln\left(P(i|\beta,C_{n})\right) \right] \label{Eq:Divergence} \\
&=& \sum_{n}\sum_{D_{n}\in C_{n}} \sum_{i\in D_{n}} \pi(D_{n}|\beta^{*}) \cdot P\left(i|\beta^{*},D_{n}\right)\cdot  \left[ ln(\pi(D_{n}|i)) -  ln\left(\frac{\sum_{j \in D_{n}} exp(V_{jn}+ln(\pi(D_{n}|j)))}{\sum_{j \in C_{n}}exp(V_{jn})}\right)\right] \label{Eq:Divergence_diff} \nonumber\\
&=& \sum_{n}\sum_{D_{n}\in C_{n}} \sum_{i\in D_{n}} \pi(D_{n}|\beta^{*}) \cdot P\left(i|\beta^{*},D_{n}\right)\cdot  \left[ ln(\pi(D_{n}|i)) -  ln\left(\sum_{j \in D_{n}} P(j|\beta,C_{n})\pi(D_{n}|j)\right)\right] \\
&=& \sum_{n}\sum_{D_{n}\in C_{n}} \sum_{i\in D_{n}} \pi(D_{n}|\beta^{*}) \cdot P\left(i|\beta^{*},D_{n}\right)\cdot  \left[ ln(\pi(D_{n}|i)) -  ln\left(\sum_{j \in C_{n}} P(j|\beta,C_{n})\pi(D_{n}|j)\right)\right] \\
&=& \sum_{n}\sum_{D_{n}\in C_{n}} \sum_{i\in D_{n}} \pi(D_{n}|\beta^{*}) \cdot P\left(i|\beta^{*},D_{n}\right)\cdot  \left[ln(\pi(D_{n}|i)) - ln\left(\pi(D_{n}|\beta)\right)\right] 
 \label{Eq:Divergence_end}
\end{eqnarray}   

The information divergence between $LL^{+}$ and $LL$ comprises two parts. First, $ln(\pi(D_{n}|i))$ captures the impact of \citet{mcfadden1978modeling}'s correction factor on the numerator of the MNL choice probability. As discussed in Section \ref{Subsec:MNL_correction}, this part is independent of $\beta$ and only scales the quasi log-likelihood down but is unrelated to the potential bias in the parameter estimates. It merely reduces the quasi (log-)likelihood to account for the fact that sampling of alternatives overestimates the MNL choice probability. The sign of the first part is negative. The second part, and again referring to Section \ref{Subsec:MNL_correction}, $ln(\pi(D_{n}|\beta))$ relates to the differences between the denominator of the quasi and 'true' MNL choice probability. This is the part where the potential bias in $\beta$ is induced by the specific sampling protocol. The sign of the second part is also negative. The sign of the information divergence thus depends on the relative size of $ln(\pi(D_{n}|j))$ and $ln(\pi(D_{n}|\beta))$. Only for uniform conditioning, we can guarantee that the information divergence is positive. In this case, $ln(\pi(D_{n}|i)) - ln\left(\pi(D_{n}|\beta)\right)$ reduces to $ln\left(\frac{\sum_{j \in C_{n}} exp(V_{jn})}{\sum_{j \in D_{n}} exp(V_{jn})}\right)>0$.  

We rewrite Eq.~\eqref{Eq:Divergence_end} using $\pi(D_{n}|\beta^{*})\cdot P(i|\beta^{*},D_{n}) = \pi(D_{n}|i)\cdot P(i|\beta^{*},C_{n})$ such that:

\begin{eqnarray}  
    \mathbb{E}\left(LL^{+}-LL\right)&=&\sum_{n}\sum_{D_{n}\in C_{n}} \sum_{i\in D_{n}} \pi(D_{n}|i)\cdot P(i|\beta^{*},C_{n}) \cdot ln(\pi(D_{n}|i)) - \sum_{n}\sum_{D_{n}\in C_{n}} \pi(D_{n}|\beta^{*}) ln\left(\pi(D_{n}|\beta)\right) \nonumber \\
    &=&\sum_{n}\sum_{D_{n}\in C_{n}} \sum_{i\in C_{n}} \pi(D_{n}|i)\cdot P(i|\beta^{*},C_{n})  \cdot ln(\pi(D_{n}|i)) - \sum_{n}\sum_{D_{n}\in C_{n}} \pi(D_{n}|\beta^{*}) ln\left(\pi(D_{n}|\beta)\right) \nonumber\\    
    &=&\sum_{n}\sum_{i \in C_{n}} P(i|\beta^{*},C_{n}) \sum_{D_{n}\in C_{n}} \pi(D_{n}|i) \cdot ln(\pi(D_{n}|i))-\sum_{n}\sum_{D_{n}\in C_{n}} \pi(D_{n}|\beta^{*})ln(\pi(D_{n}|\beta)) \nonumber 
\end{eqnarray}

Only the term $\sum_{D_{n}\in C_{n}} \pi(D_{n}|\beta^{*}) \cdot ln(\pi(D_{n}|\beta))$ depends on $\beta$. Since $\sum_{D_{n}\in C_{n}}\pi(D_{n}|\beta)=1$, maximum entropy arises at the true parameter $\beta=\beta^{*}$ and $\mathbb{E}\left(LL^{+}-LL\right)$ is minimised with respect to $\beta$. This result, however, only applies when $\mathbb{E}\left(LL^{+}-LL\right)>0$. This includes uniform conditioning and a limited but unknown set of sampling protocols satisfying positive conditioning. \citet{mcfadden1978modeling}'s correction factor thus not only results in consistent parameter estimates but this consistent parameter estimate also minimises the expected information divergence between the quasi and 'true' log-likelihood for certain sampling protocols assuming that the data generating process is MNL. This supports \citet{keane2012estimation}'s statement that uniform conditioning shifts up the expected log-likelihood, but does not alter where it is maximised.

The fact that \citet{mcfadden1978modeling}'s correction factor does not minimise the expected information divergence for all sampling protocols satisfying positive conditioning is not a cause for concern. Namely, all sampling protocols satisfying positive conditioning maximise $\sum_{D_{n}\in C_{n}} \pi(D_{n})^{*} \cdot ln(\pi(D_{n}|\beta))$ and thereby minimise the (positive) expected information loss with respect to the parameters of interest (i.e. consistent parameter estimates are obtained). The information divergence may only become negative because $\sum_{D_{n}\in C_{n}} \pi(D_{n}|i) \cdot ln(\pi(D_{n}|i))$ may over-correct for the over-estimation of the choice probability under the sampling of alternatives independently of $\beta$. For uniform conditioning, there is only one source of information divergence (i.e. through the denominator) which may cause bias and minimising the information divergence corresponds with minimising this expected information loss with respect to $\beta$. 

\section{Minimum expected information loss for Bayesian MMNL models using data augmentation}
\label{App:EDKL}
When applying data augmentation the joint posterior density for the MMNL model can be described by
\begin{equation}
    p(\beta^{+},\theta|Y,C) = \frac{p(\theta) \cdot \prod_{n=1}^{N}f(\beta_{n}|\theta) \cdot \prod_{t=1}^{T} P(Y_{nt}|\beta_{n},C_{nt})}{\int_{\theta} p(\theta) \cdot \prod_{n=1}^{N}\int_{\beta_{n}}f(\beta_{n}|\theta) \cdot \prod_{t=1}^{T} P(Y_{nt}|\beta_{n},C_{nt}) d\beta_{n}d\theta}
\end{equation}

\noindent Likewise, the same approximate density under the sampling of alternatives can be described by
\begin{equation}
    p(\beta^{+},\theta|Y,D) = \frac{p(\theta) \cdot \prod_{n=1}^{N}f(\beta_{n}|\theta) \cdot \prod_{t=1}^{T} P(Y_{nt}|\beta_{n},D_{nt})}{\int_{\theta} p(\theta) \cdot \prod_{n=1}^{N}\int_{\beta_{n}}f(\beta_{n}|\theta) \cdot \prod_{t=1}^{T} P(Y_{nt}|\beta_{n},D_{nt}) d\beta_{n}d\theta}
\end{equation}

Linking back to Eqs.~\eqref{Eq:KLBASIC1}-\eqref{Eq:KLBASIC2}, the $D_{KL}$ measure summarising the information loss due to approximating the true distribution, comprises a part relating to the loss of information with respect to the parameters of interest and a part relating to the loss in model fit. 

\begin{eqnarray}
    D_{KL}&=&\int_{\beta^{+}}\int_\theta p(\beta^{+},\theta|Y,C) \cdot ln\left(\frac{p(\theta) \cdot \prod_{n=1}^{N}f(\beta_{n}|\theta) \cdot \prod_{t=1}^{T} P(Y_{nt}|\beta_{n},C_{nt})}{p(\theta) \cdot \prod_{n=1}^{N}f(\beta_{n}|\theta) \cdot \prod_{t=1}^{T} P(Y_{nt}|\beta_{n},D_{nt})}\right)d\theta d\beta^{+} \nonumber \\ 
    &+& ln\left(\frac{\int_{\theta} p(\theta) \cdot \prod_{n=1}^{N}\int_{\beta_{n}}f(\beta_{n}|\theta) \cdot \prod_{t=1}^{T} P(Y_{nt}|\beta_{n},D_{nt}) d\beta_{n}d\theta}{\int_{\theta} p(\theta) \cdot \prod_{n=1}^{N}\int_{\beta_{n}}f(\beta_{n}|\theta) \cdot \prod_{t=1}^{T} P(Y_{nt}|\beta_{n},C_{nt}) d\beta_{n}d\theta}\right)
\end{eqnarray}

 Under the assumption of identical priors $p(\theta)$ and mixing density $f(\beta_{n}|\theta)$ between the true and sampled model, the $D_{KL}$ measure reduces to: 

\begin{equation}
    D_{KL}=\int_{\beta^{+}}\int_\theta p(\beta^{+},\theta|Y,C) \cdot ln\left(\frac{\prod_{n=1}^{N}\prod_{t=1}^{T} P(Y_{nt}|\beta_{n},C_{nt})}{\prod_{n=1}^{N}\prod_{t=1}^{T} P(Y_{nt}|\beta_{n},D_{nt})}\right)d\theta d\beta^{+} + ln (A) 
\end{equation}

Consistent with Section \ref{Sec:BayesMNL}, only the first part of the $D_{KL}$ measure is of interest for the purposes of estimation. Following Eqs.~\eqref{Eq:InfDiv_Bayes1}-\eqref{Eq:InfDiv_Bayes2}, we can rewrite the expression inside the integral of the first part to:

\begin{equation}
    ln\left(\frac{\prod_{n=1}^{N}\prod_{t=1}^{T} P(Y_{nt}|\beta_{n},C_{nt})}{\prod_{n=1}^{N}\prod_{t=1}^{T} P(Y_{nt}|\beta_{n},D_{nt})}\right)=\sum_{n=1}^{N}\sum_{t=1}^{T} ln(\pi(D_{nt}|\beta_{n}))-ln(\pi(D_{nt}|i)
\end{equation}

Similar to Section \ref{Sec:BayesMNL}, we can recognise that $\sum_{n=1}^{N}\sum_{t=1}^{T}ln(\pi(D_{nt}|i))$ operates as a scalar and is independent of $\beta_{n}$ and $\theta$, and therefore is unrelated to the information loss with respect to these parameters of interest. The term can therefore be disregarded. 
The information loss with respect to $\beta_{n}$ and $\theta$ - as represented by $\sum_{n=1}^{N}\sum_{t=1}^{T} ln(\pi(D_{nt}|\beta_{n}))$  is negative for all sampling protocols satisfying either uniform or positive conditioning. We aim to minimise the expected information loss in Eq.~\eqref{Eq:MMNL_infDiv} with respect to $\beta_{n}$ and $\theta$. Define the joint probability of observing $\theta$, $\beta_{n}$, the choice for alternative $i$ and sampled choice set $D_{nt}$ by $p(\theta,\beta_{n},i,D_{nt})=p(\theta)f(\beta_{n}|\theta)P(i|\beta_{n},D_{nt})\pi(D_{nt}|\beta_{n})$ such that:

\begin{equation}    
\label{Eq:MMNL_infDiv}
    \mathbb{E}\left(\sum_{n=1}^{N}\sum_{t=1}^{T} ln(\pi(D_{nt}))\right)= \int_{\theta} p(\theta) \sum_{n=1}^{N} \int_{\beta_{n}}f(\beta_{n}|\theta) \sum_{t=1}^{T} \sum_{D_{nt}\in C_{nt}} \pi(D_{nt}|\beta_{n}) ln(\pi(D_{nt}|\beta_{n})) d\beta_{n}d\theta   
\end{equation}

Since the entropy $\sum_{D_{nt} \in C_{nt}} \pi(D_{nt}) ln(\pi(D_{nt}))$ is maximised at \citet{mcfadden1978modeling}'s correction term for every value of $\theta$ and $\beta_{n}$, \citet{mcfadden1978modeling}'s correction factor thus minimises the expected information loss across the entire posterior for $\beta_{n}$ and $\theta$, not just at the 'true' parameters. This result applies to any sampling protocol satisfying either a uniform or positive conditioning under the assumption that the data generating process is MMNL, \citet{mcfadden1978modeling}'s result for MNL transfers to the Bayesian MMNL when data augmentation is implemented. 
\end{document}